\documentclass{article}
\usepackage{tabularray}
\usepackage{amssymb}
\usepackage{graphicx} % Required for inserting images
\usepackage[utf8]{inputenc}
\usepackage{geometry}
\usepackage{color}
\usepackage[table,xcdraw]{xcolor}
\usepackage[T1]{fontenc}
\usepackage[utf8]{inputenc}
\usepackage{graphicx}
\usepackage{lmodern}
\usepackage[english]{babel}
\usepackage{graphicx}
\usepackage{float}
\usepackage{siunitx}
\usepackage{multirow}
\usepackage{lscape}
\usepackage{booktabs}
\usepackage{subcaption}
\usepackage{amsmath}
\usepackage{rotating}
\usepackage{longtable}
\usepackage{authblk}
\usepackage{fullpage}
\usepackage[numbers,sort&compress]{natbib}
\usepackage[colorlinks,
            linkcolor=black,
            anchorcolor=black,
            citecolor=black
            ]{hyperref}
\usepackage{acronym} 
\graphicspath{{figures/}}
\newcommand{\leff}{l_{\text{eff}}}
\newcommand{\aeff}{A_{\text{eff}}}
\newcommand{\aann}{A_{\text{ann}}}
\DeclareSIUnit\mmHg{mmHg}
\title{Cardiovascular function changes following lung resection: \\ a computational model to compare afterload increase and contractility loss mechanisms
}

\begin{document}
\providecommand{\keywords}[1]
{\textbf{\text{Keywords:}}#1}
\author[1,2]{Shiting Huang}
\author[3]{Sanjay Pant}
\author[2,4]{Sean McGinty}
\author[5,6]{Richard Good}
\author[5,7]{Ben Shelley}
\author[1,2]{Ankush Aggarwal\thanks{Corresponding author: Ankush.Aggarwal@glasgow.ac.uk}}

\affil[1]{Division of Infrastructure \& Environment, James Watt School of Engineering, University of Glasgow, Glasgow, UK}
\affil[2]{Glasgow Computational Engineering Centre, University of Glasgow, Glasgow, UK}
\affil[3]{CardioLume, Swansea, UK}
\affil[4]{Division of Biomedical Engineering, James Watt School of Engineering, University of Glasgow, Glasgow, UK}
\affil[5]{Golden Jubilee National Hospital, Glasgow, UK}
\affil[6]{British Heart Foundation Glasgow Cardiovascular Research Centre, University of Glasgow, Glasgow, UK}
\affil[7]{Anaesthesia, Perioperative Medicine and Critical Care Research Group, University of Glasgow, Glasgow, UK}

\maketitle

\section*{Abstract}
Functional limitation after lung resection surgery has been consistently documented in clinical studies, and right ventricle (RV) dysfunction has been hypothesized as a contributing reason. However, the mechanisms of RV dysfunction after lung resection remain unclear, particularly whether change in afterload or contractility is the main cause. This study is the first to employ a lumped parameter model to simulate the effects of lung resection. The implementation of a computational model allowed us to isolate certain mechanisms that are difficult to perform clinically. Specifically, two mechanisms were compared: afterload increase and RV contractility loss. Furthermore, our rigorous approach included local and global sensitivity analyses to evaluate the effect of parameters on our results, both individually and collectively. Our results demonstrate that contractility and afterload exhibited consistent trends across various pressure and volume conditions, pulmonary artery systolic pressure, pulmonary artery diastolic pressure, and right ventricular systolic pressure showed opposite variations. The results show that post-operative RV dysfunction may result from a combination of RV contractility loss and afterload increase. Further exploration and refinement of this first computational model presented herein will help us predict RV dysfunction after lung resection and pave the way towards improving outcomes for lung cancer patients.

\keywords{ lung resection; cardiovascular; lumped parameter modelling; right ventricle; RV dysfunction; contractility; numerical simulation; afterload}

\section{Introduction}\label{intro}
Lung cancer ranks as the second leading cause of cancer mortality globally and is the third most commonly diagnosed cancer \cite{Travis_2011, Vainshelboim_Fox_Saute_Sagie_Yehoshua_Fuks_Schneer_Kramer_2015}, and it exerts a considerable impact on public health. Among the therapeutic options, lung resection surgery is recognized as an effective treatment for lung cancer. According to guidelines issued by the National Institute for Health and Care Excellence (NICE), in suitable cases, lung resection is recommended as a prioritized intervention \cite{niceOverviewLung}. Lung resection surgery varies based on the specific location and extent of the disease within the lung. Anatomically, there are two lungs, one on the left and one of the right side. The lungs are usually divided into five lobes: left lung has two lobes and the right lung has three lobes. Furthermore, these lobes can be categorized into a total of nineteen segments, with the left lung comprising nine segments and the right lung containing ten. 
Consequently, the lung resection surgeries are categorized into lobectomy (removal of one or two lobes) and pneumonectomy (removal of the whole left or right lung).

Although lung resection is recognized as an effective treatment for lung cancer, it is not free of consequences. Patients may have significant issues after surgery, including disabling dyspnea (shortness of breath) and decreased functional capacity \cite{McCall_Arthur_Glass_Corcoran_Kirk_Macfie_Payne_Johnson_Kinsella_Shelley_2019}. Historically, these post-operative issues have been attributed to diminished lung function (as a result of a part of the lung physically removed); however, subsequent studies have demonstrated only a weak association with lung function tests \cite{Pelletier_Lapointe_LeBlanc_1990, R._Larsen_Svendsen_Milman_Brenøe_Petersen_1997}. An alternative hypothesis suggests that cardiac function may contribute to these post-operative problems due to the interaction between the heart and pulmonary circulation \cite{McCall_Arthur_Glass_Corcoran_Kirk_Macfie_Payne_Johnson_Kinsella_Shelley_2019, Vainshelboim_Fox_Saute_Sagie_Yehoshua_Fuks_Schneer_Kramer_2015}. This perspective is supported by observation data showing a high prevalence of right ventricular (RV) dysfunction following a lung resection surgery, most commonly quantified as a decreased RV ejection fraction (RVEF) \cite{Mageed_FaragEl-Ghonaimy_Elgamal_Hamza_2005, McCall_Arthur_Glass_Corcoran_Kirk_Macfie_Payne_Johnson_Kinsella_Shelley_2019, Elrakhawy_Alassal_Shaalan_Awad_Sayed_Saffan_2018, Okada_Ota_Okada_Matsuda_Okada_Ishii_1994, Kowalewski_Brocki_Dryjański_Kaproń_Barcikowski_1999}. Despite extensive clinical research over the past decades, the mechanisms underlying post-operative RV dysfunction remain elusive \cite{Gelzinis_Assaad_Perrino_2020}. A proven factor is the change in RV afterload after surgery \cite{Shelley_Glass_Keast_McErlane_Hughes_Lafferty_Marczin_McCall_2023}. However, it is unclear if afterload increase caused by lung resection is sufficient to fully explain the observed postoperative RV dysfunction. In previous animal studies involving operations functionally analogous to lung resection  \cite{Greyson_Xu_Cohen_G._Schwartz_1997}, RV function continued to decline persistently even after normalization of afterload. It has also been noted that afterload changes trigger inflammatory injury in the RV \cite{Watts_Marchick_Kline_2010, Watts_Gellar_Stuart_Obraztsova_Kline_2009}. Thus, an impairment of RV contractility caused by such an inflammatory injury is suspected to be partly responsible for postoperative RV dysfunction \cite{Shelley_Glass_Keast_McErlane_Hughes_Lafferty_Marczin_McCall_2023}.

Clinical measurements are often unable to isolate the effects of individual factors due to the complex interplay within the physiological systems. Thus, there remains an ongoing debate whether changes in afterload following lung resection or alterations in RV contractility are primarily responsible for RV dysfunction \cite{Reed_Dorman_Spinale_1993, Reed_Dorman_Spinale_1996, Okada_Ota_Okada_Matsuda_Okada_Ishii_1994, McCall_Arthur_Glass_Corcoran_Kirk_Macfie_Payne_Johnson_Kinsella_Shelley_2019}. In this study, we review the main clinical studies of RV dysfunction following lung resection and then describe a novel computational model to simulate the isolated effects of afterload and RV contractility on the cardiovascular system following lung resection surgery. Such a modelling framework has the potential to be instrumental in elucidating the specific mechanisms underlying functional limitations in lung resection surgery patients. A deeper understanding of these mechanisms is clinically crucial, as it will guide the development of targeted preventative and therapeutic strategies to effectively minimize these complications.

\section{Review of Clinical Literature}
Clinical research on RV function following lung resection spans approximately eight decades. The earliest documented study was conducted by \citet{Rams_Harrison_Fry_Moulder_Adams_1962}. Following a period of limited investigation, the field advanced in the 1990s through the seminal works of Reed et al., Okada et al., and Nishimura et al., who extensively investigated RV function and post-operative exercise capacity \cite{Reed_Spinale_Crawford_1992,Reed_Dorman_Spinale_1993,Okada_Ota_Okada_Matsuda_Okada_Ishii_1994,Okada_Okada_Ishii_Yamashita_Sugimoto_Okada_Yamagishi_Yamashita_Matsuda_1996,Nishimura_Haniuda_Morimoto_Kubo_1993}. Notable contributions after 2000s include the comprehensive analysis by \citet{Elrakhawy_Alassal_Shaalan_Awad_Sayed_Saffan_2018}, featuring the largest sample size to date, and the pioneering works by \citet{Shelley_Glass_Keast_McErlane_Hughes_Lafferty_Marczin_McCall_2023} and \citet{Glass_McCall_Arthur_Mangion_Shelley_2023} exploring the effect of wave reflection in the pulmonary circulation after the surgery.

In all of these studies, RV dysfunction is a well-documented complication following lung resection, but the mechanism remains unclear. Many studies have observed a short-term postoperative rise in pulmonary vascular resistance (PVR) and pulmonary arterial pressure (PAP), indicating increased afterload. However, \citet{Shelley_Glass_Keast_McErlane_Hughes_Lafferty_Marczin_McCall_2023} emphasized that these changes reflect static afterload and do not include pulsatile components, such as pulse wave reflection, compliance, and inertia, which are essential for understanding the true afterload. Consequently, further exploration of pulsatile components is necessary to better understand their relationship with RV dysfunction. For example, more recent studies \cite{McCall_Arthur_Glass_Corcoran_Kirk_Macfie_Payne_Johnson_Kinsella_Shelley_2019, Glass_McCall_Arthur_Mangion_Shelley_2023} have confirmed the relationship between increased RV afterload and dysfunction by identifying wave reflections in the pulmonary vasculature of the operated lung, demonstrating how altered hemodynamics contribute to RV impairment. In contrast, there are few studies on changes in RV contractility after lung resection.

\subsection{Cardiovascular Measurements and Findings}

Since the present study focusses on the cardiovascular system and its function change after lung resection, existing clinical literature on pressure- and volume-related quantities and how these changes post-operatively are reviewed in detail next. The most relevant clinical findings of function change are summarised in Tables \ref{RVdetail} and \ref{PAdetail}. To aid understanding,  definitions of some of the commonly physiological abbreviations are listed in Appendix \ref{clinical definition}.
\subsubsection{RV volume and RVEF}
Multiple studies have documented increases in post-op RV volumetric measurements following lung resection. \citet{Reed_Spinale_Crawford_1992,Reed_Dorman_Spinale_1993}, \citet{Kowalewski_Brocki_Dryjański_Kaproń_Barcikowski_1999}, \citet{McCall_Arthur_Glass_Corcoran_Kirk_Macfie_Payne_Johnson_Kinsella_Shelley_2019} and \citet{Elrakhawy_Alassal_Shaalan_Awad_Sayed_Saffan_2018} reported elevations in RV end-diastolic volume index (RVEDVI). \citet{Kowalewski_Brocki_Dryjański_Kaproń_Barcikowski_1999} and \citet{McCall_Arthur_Glass_Corcoran_Kirk_Macfie_Payne_Johnson_Kinsella_Shelley_2019} also noted RV end-systolic volume index (RVESVI) increases. 
These findings were further corroborated by \citet{Rauch_Marinova_Schild_Strunk_2017} using CT measurements of the RV diameter.

RV stroke volume index (RVSVI), defined as the difference between RVEDVI and RVESVI, and RVEF, defined as RVSVI/RVEDVI, are two volume-derived measures. RVEF is the widely accepted gold indicator of the RV function. 
 Multiple studies have consistently observed post-operative RVEF decline \cite{McCall_Arthur_Glass_Corcoran_Kirk_Macfie_Payne_Johnson_Kinsella_Shelley_2019, Elrakhawy_Alassal_Shaalan_Awad_Sayed_Saffan_2018, Reed_Dorman_Spinale_1996, Reed_Dorman_Spinale_1993, Reed_Spinale_Crawford_1992,Okada_Ota_Okada_Matsuda_Okada_Ishii_1994, Kowalewski_Brocki_Dryjański_Kaproń_Barcikowski_1999}, providing substantial evidence for post-operative RV dysfunction \cite{Shelley_Glass_Keast_McErlane_Hughes_Lafferty_Marczin_McCall_2023}.
\subsubsection{RV pressure}
 Regarding RV pressure, both \citet{Amar_Burt_Roistacher_Reinsel_Ginsberg_Wilson_1996} and \citet{Mandal_Dutta_Kumar_Kumar_Ganesan_Bhat_2017} observed post-operative increases in RV systolic pressure (RVSP). Specifically, \citet{Amar_Burt_Roistacher_Reinsel_Ginsberg_Wilson_1996} reported mild RVSP increases of 4.16\% and 3.85\% on post-operative day 1 following lobectomy and pneumonectomy. However, these changes were not statistically significant ($p$-value > 0.05). Similarly, \citet{Mandal_Dutta_Kumar_Kumar_Ganesan_Bhat_2017} observed larger RVSP increases of 86.16\% and 123\% on post-operative day 2 for lobectomy and pneumonectomy. Despite the larger increase, the $p$-values were still greater than 0.05, indicating no statistical significance.

\subsubsection{Pulmonary arterial pressure and pulmonary vascular resistance}
\citet{Reed_Spinale_Crawford_1992,Reed_Dorman_Spinale_1993} documented increases in mean pulmonary artery (PA) pressure (mPAP), PA diastolic pressure (PADP), and PA systolic pressure (PASP), postoperatively. Similarly, \citet{Elrakhawy_Alassal_Shaalan_Awad_Sayed_Saffan_2018} and \citet{Nishimura_Haniuda_Morimoto_Kubo_1993}  reported elevated mPAP. Notably, \citet{Okada_Ota_Okada_Matsuda_Okada_Ishii_1994} found different outcomes; they observed no significant changes in mPAP and PASP. 
Pulmonary vascular resistance (PVR), a critical component of RV afterload, is reported by most studies to increase following surgery \cite{Rauch_Marinova_Schild_Strunk_2017, Reed_Spinale_Crawford_1992, Nishimura_Haniuda_Morimoto_Kubo_1993}. 

\subsubsection{Left Ventricular Response}
While most investigations have centered on RV-pulmonary interactions, limited studies have examined the left ventricular (LV) response to lung resection. \citet{Okada_Ota_Okada_Matsuda_Okada_Ishii_1994} reported stable LV function indices, such as the arterial pressure. \citet{Mandal_Dutta_Kumar_Kumar_Ganesan_Bhat_2017} similarly observed maintained LV ejection fraction and strain, post-operatively.

In summary, clinical evidence consistently demonstrates post-operative increases in RV volume, accompanied by a decreased RVEF. Pulmonary arterial pressure and resistance typically rise, while LV function remains largely unchanged.

\definecolor{Sunglo}{rgb}{0.89,0.392,0.392}
\definecolor{SeaPink}{rgb}{0.917,0.6,0.6}
\definecolor{ChetwodeBlue}{rgb}{0.494,0.635,0.874}
\definecolor{CatskillWhite}{rgb}{0.933,0.949,0.964}
\definecolor{TropicalBlue}{rgb}{0.854,0.917,0.98}
\definecolor{ChetwodeBlue1}{rgb}{0.521,0.65,0.87}
\begin{table}
\centering
\caption{The trend change of most relevant volume measurements in RV from the literature. Some studies do not use the body surface area to normalize (index) the volume \cite{Reed_Spinale_Crawford_1992, McCall_Arthur_Glass_Corcoran_Kirk_Macfie_Payne_Johnson_Kinsella_Shelley_2019}; statistically significant quantities are indicated by $\dagger$. Lob represents lobectomy, and Pnx is pneumonectomy.}
\label{RVdetail}
\resizebox{\textwidth}{!}{%
\begin{tblr}{
  cells = {c},
  cell{4}{1} = {r=2}{},
  cell{7}{1} = {r=2}{},
  hline{1-4,6-7,9} = {-}{},
}
Source                                                                                 & Cohort size & Resection Type & $\Delta$RVEDVI            & $\Delta$RVESVI             & $\Delta$RVSVI             & $\Delta$RVEF               \\
\citet{Reed_Spinale_Crawford_1992}                                                          & 15          & Lob           & $+$12.7\%            &                    &                   & $-$8.9\%$^\dagger$  \\
\citet{Reed_Dorman_Spinale_1993}                                                          & 10          & Lob+Pnx        &                   & $+$37.3\%$^\dagger$  & $+$15.8\%$^\dagger$ & $-$7.1\%            \\
\citet{Kowalewski_Brocki_Dryjański_Kaproń_Barcikowski_1999}                             & 9           & Lob           & $+$3.4\%            & $+$5.5\%             & $+$1.0\%            & $-$4.3\%            \\
                                                                                       & 22          & Pnx           & $+$20.8\%$^\dagger$ & $+$42.7\%$^\dagger$  & $-$2.2\%           & $-$18.8\%$^\dagger$ \\
\citet{McCall_Arthur_Glass_Corcoran_Kirk_Macfie_Payne_Johnson_Kinsella_Shelley_2019} & 26–22       & Lob+Pnx        & $+$5.7\%$^\dagger$  & $+$14.7\%$^\dagger$ & $-$3.4\%$^\dagger$ & $-$11.1\%$^\dagger$ \\
\citet{Elrakhawy_Alassal_Shaalan_Awad_Sayed_Saffan_2018}                                & 142         & Lob           & $+$18.7\%$^\dagger$ &                    &                   & $-$15.5\%$^\dagger$ \\
                                                                                       & 36          & Pnx           & $+$37.7\%$^\dagger$ &                    &                   & $-$21.6\%$^\dagger$ 
\end{tblr}
}
\end{table}

\begin{table}
\centering
\caption{The trend change of most relevant pulmonary pressure measurements from the literature.  Statistically significant results are indicated with $\dagger$. Lob represents lobectomy, and Pnx is pneumonectomy.}
\label{PAdetail}
\resizebox{\textwidth}{!}{%
\begin{tblr}{
  cells = {c},
  cell{4}{1} = {r=2}{},
  hline{1-4,6-8} = {-}{},
}
Source                                                                          & Cohort size & Resection Type          & $\Delta$mPAP               & $\Delta$PASP    & $\Delta$PADP    & $\Delta$PVR                 \\
\citet{Reed_Spinale_Crawford_1992}                      & 15          & Lob                    & $+$10.0\%               & $+$9.7\%  & $+$23.1\% & $+$21.6\%             \\
\citet{Reed_Dorman_Spinale_1993}                           & 10          & Lob+Pnx & $+$15.8\%            & $+$6.9\%   &         &                     \\
\citet{Elrakhawy_Alassal_Shaalan_Awad_Sayed_Saffan_2018}& 142         & Lob                    & $+$51.5\%$^\dagger$ &         &         & $+$82.3\%$^\dagger$  \\
                                                                                & 36          & Pnx                    & $+$98.4\%$^\dagger$  &         &         & $+$107.7\%$^\dagger$ \\
\citet{Okada_Ota_Okada_Matsuda_Okada_Ishii_1994}       & 20          & Lob+Pnx & $-$6.3\%            & $-$3.9\% &         & $-$12.0\%            \\
\citet{Nishimura_Haniuda_Morimoto_Kubo_1993}               & 9           & Lob                    & $+$13.7\%             &         &         & $+$33.0\%                
\end{tblr}
}
\end{table}

\section{Computational Modeling Methodology}\label{method}
Although the hypothesis that afterload has a correlation with post-operative RV dysfunction is supported, it remains unclear whether RV contractility is also related to RV dysfunction after the lung resection. In addition, there is uncertainty about the relative influence of each of these parameters on the cardiovascular system, largely due to the difficulty of independently altering them in clinical research. Instead, employing computational models to simulate the responses to lung resection surgery may provide unique insights into the underlying mechanisms of RV dysfunction. While there is a large amount of literature on the computational modeling of cardiovascular system, to the best of our knowledge, no computational/mathematical modeling of the effect of lung resection on cardiovascular system has been reported. 

In the field of computational modeling of the cardiovascular system, existing methodology can generally be categorized into two distinct approaches: 1) the lumped parameter model (0D model), which employs a hydraulic-electrical analogy to simulate the global cardiovascular response; and 2) the distributed parameter model (1D, 2D, or 3D model), which aims to capture the spatially distributed responses within the system \cite{Korakianitis_Shi_2006, Shi_Lawford_Hose_2011}. For this study, we opted to utilize the lumped parameter (0D) model due to the challenges associated with constructing a complete representation of the human vascular network, which extends over 10,000 km, using the distributed model \cite{Mescher_Junqueira_2018}. 

\subsection{Model Structure}\label{model}
\begin{figure}[h!]
    \centering\includegraphics[width=1\linewidth]{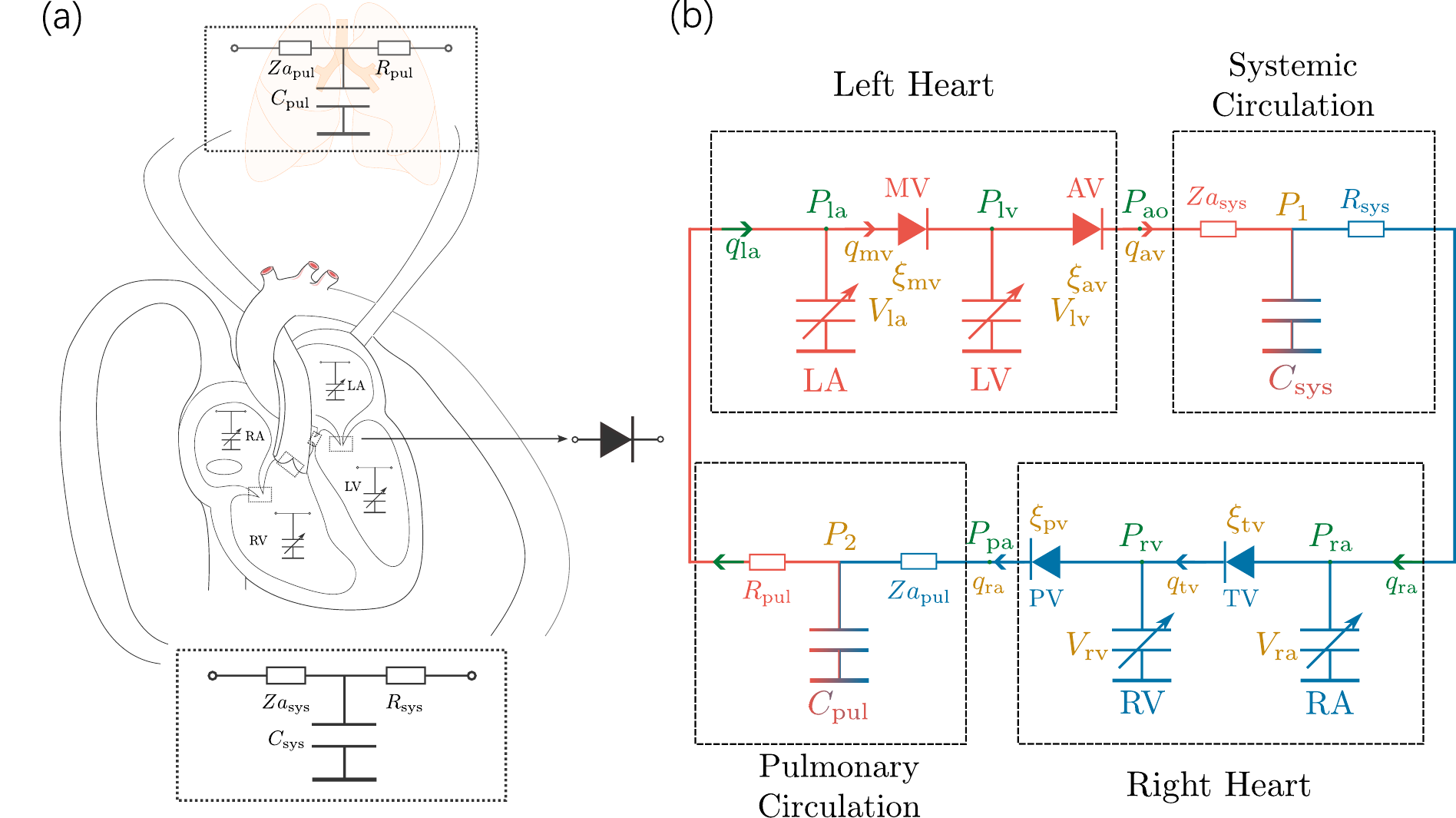}
    \caption{(a) Physiological circulatory system; (b) Computational model structure corresponding to (a). Here, arrows represent the blood flow direction, red represents oxygenated blood, blue represents deoxygenated blood. 
LA: left atrium; LV: left ventricle; RA: right atrium; RV: right ventricle; MV: mitral valve; AV: aortic valve; TV: tricuspid valve; PV: pulmonary valve; subscripts $_{la}$, $_{lv}$, $_{ra}$, $_{rv}$, $_{mv}$, $_{av}$, $_{pv}$, $_{tv}$ also represent the same meaning of their upper case counterparts. $P$: pressure; $q$: flow; $Za$: input impedance of circulation; $R$: peripheral resistance of circulation; $C$: capacitance of circulation. Variables colored yellow are the state variables used when solving the model dynamics.}
    \label{model structure}
\end{figure}

The proposed 0D closed-loop model is based on a hydraulic-electric analogy, where the current
represents blood flow and voltage represents pressure \cite{Olufsen_Nadim}. The model is composed of three main components: four heart chambers (time-variant elastance), four valves (diodes), and two circulations (resistors and capacitors) (Figure \ref{model structure}). Each of these are described next.

\begin{figure}
    \centering
    \includegraphics[width=0.7\linewidth]{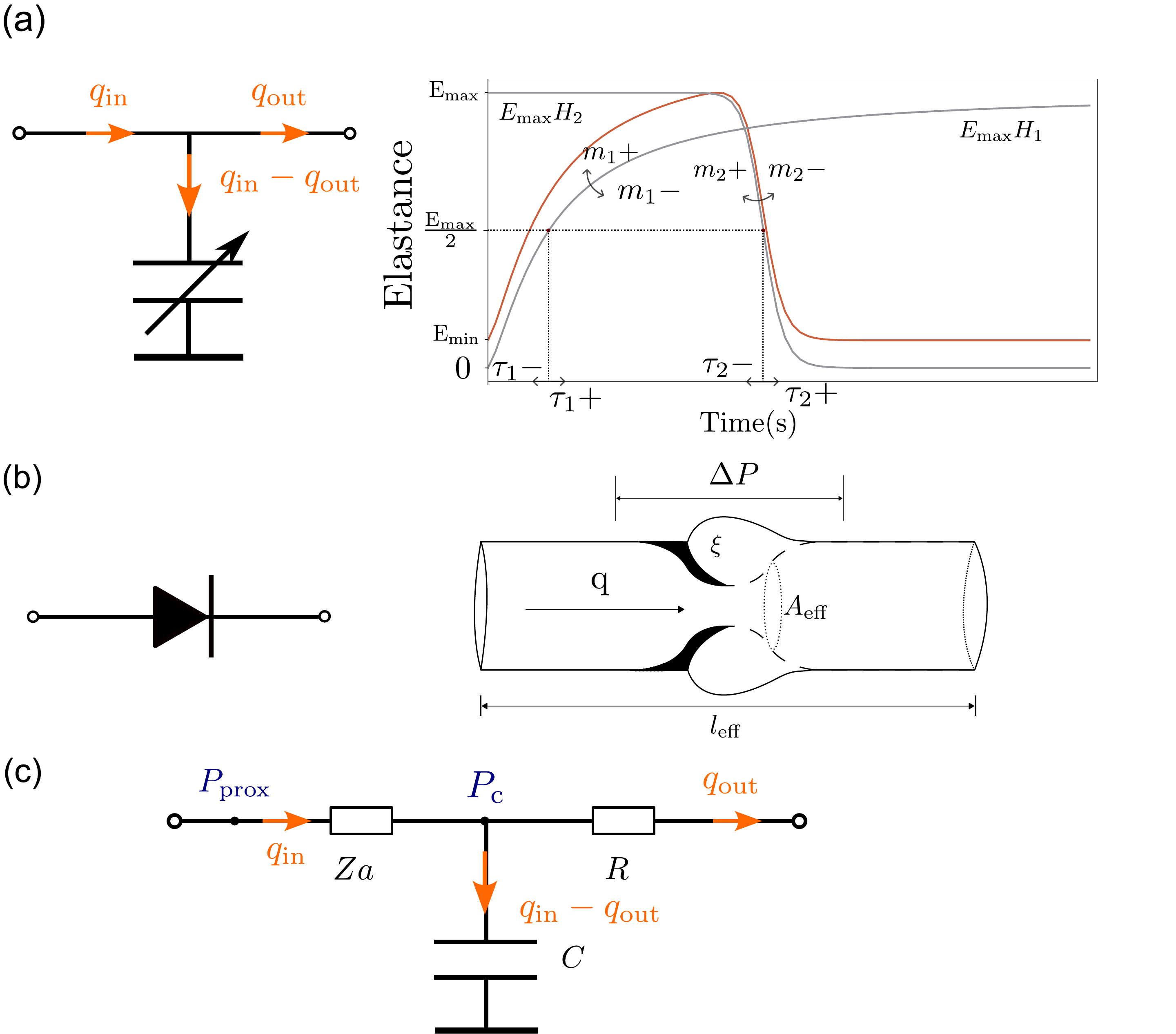}
    \caption{Representation of each component in the lumped parameter model: (a) The heart chamber components utilize a time-varying elastance function. The parameters $m_{1}$ and $m_{2}$ control the slopes of $H_{1}$ and $H_{2}$, respectively, while $\tau_{1}$ and $\tau_{2}$ determine the timing at which $E_{\max}$ reaches 50\%; (b) The valve component considers turbulence and inertance effects through the valve's open/close state parameter $\xi$, effective length $l_\text{eff}$, and pressure drop $\Delta P$. (c) The blood flow into the circulation will be split in two, one part will flow into the capacitor as blood volume in the vessels, and the other part will continue to flow in the direction of the resistor to the rest of the circulatory system
}
    \label{model_component}
\end{figure}

\paragraph{Heart chamber} The heart chamber model adopts a time-variant elastance function \cite{Suga_Sagawa_Shoukas_1973, Suga_Sagawa_1974}. In this model, the elastance function $E(t)$ is specified as an input and represents the active contraction of the heart chamber. The contraction and relaxation of the heart therefore can be controlled by $E(t)$, and, as a result, produce varying pressure (Figure \ref{model_component}(a)). The pressure generated by the heart chamber serves as the driving force for blood flow throughout the cardiovascular system. The pressure in the heart chamber $P(t)$ is related to its volume $V(t)$ via the elastance function as 
\begin{equation}\label{heart chamber}
    P(t) = E(t)(V(t) - V_{0}),
\end{equation}
where $V_{0}$ is the intercept volume (the volume at zero pressure) of the chamber. The function \(E(t)\) is periodic with period \(T\), and it is defined as:
\begin{equation}
\begin{aligned}
         E(t) &= \frac{E_{\max}-E_{\min}}{\max_t\left( H_1(\bar{t}) H_2(\bar{t})\right)} H_{1}(\bar{t}) H_{2}(\bar{t}) +E_{\min},\\
              \bar{t}  &=
\begin{cases} 
%t \bmod T, & \text{if } t \geq 0 \text{ (ventricle)} \\
(t - T_{d}) \bmod T, & \text{if } t \geq T_d \\%\text{ (atrium)}\\
0, & \text{if } t < T_{d} %\text{ (atrium)}
\end{cases} \\
         H_{1}(\bar{t}) &= \frac{\left(\frac{\bar{t}}{\tau_{1}}\right)^{m_{1}}}{1+\left(\frac{\bar{t}}{\tau_{1}}\right)^{m_{1}}},\\
         H_{2}(\bar{t}) &= \frac{1}{1+\left(\frac{\bar{t}}{\tau_{2}}\right)^{m_{2}}},
\end{aligned}
\end{equation}
where \(E_{\min}\) and \(E_{\max}\) are the minimum and maximum values of elastance, respectively. The use of the modulo operation \((t \bmod T)\) ensures that the adjusted time $\bar{t}$ falls within the range of a single period \([0, T)\), effectively capturing the periodic behavior of \(E(t)\). $\tau_{1}$ and $\tau_{2}$ are the parameters that control duration of cardiac activation, and $m_{1}$ and $m_{2}$ are the parameters that control the slopes of $H_{1}$ and $H_{2}$ \cite{Mynard_Davidson_Penny_Smolich_2012,Stergiopulos_Meister_Westerhof_1996} (Figure~\ref{model_component}a). Physiologically, the atria and ventricles do not contract at the same time, which is modelled using a time lag parameter $T_{d}$.

By mass conservation, the change of heart chamber volume is the imbalance between blood flow into and out of the heart chamber (Figure \ref{model_component}a) \cite{Chung_Niranjan_Clark_Bidani_Johnston_Zwischenberger_Traber_1997}. That is,
\begin{equation}
    \frac{dV}{dt} = q_{\text{in}} - q_{\text{out}},
\end{equation}
where $q_{\text{in}}$ and $q_{\text{out}}$ are the instantaneous flow rates in and out of the heart chamber. All heart chamber parameter values used in this work are provided in Table \ref{heart_chamber_parameter}.\\
\begin{table}[htbp]
\centering
\caption{Parameter values of heart chambers (taken from \cite{Mynard_Davidson_Penny_Smolich_2012}), with 
$T=0.8$ s representing the time period of the cardiac cycle.}
\label{heart_chamber_parameter}
\resizebox{\textwidth}{!}{
\begin{tabular}{@{}cccccccc@{}}
\toprule
\textbf{Heart Chamber} & \multicolumn{1}{c}{\textbf{Left Atrium}} & \multicolumn{1}{c}{\textbf{Right Atrium}} & \multicolumn{1}{c}{\textbf{Left Ventricle}} & \multicolumn{1}{c}{\textbf{Right Ventricle}} & \textbf{Units} \\
\cmidrule(r){2-2} \cmidrule(lr){3-3} \cmidrule(lr){4-4} \cmidrule(l){5-5}
\textbf{Parameter} & {\textbf{Value}} & {\textbf{Value}} & {\textbf{Value}} & {\textbf{Value}} & \\
\midrule
$\tau_1$ & 0.110$T$ & 0.110$T$ & 0.269$T$ & 0.269$T$ & s \\
$\tau_2$ & 0.180$T$ & 0.180$T$ & 0.452$T$ & 0.452$T$ & s \\
$E_{max}$ & 0.17 & 0.15 & 3 & 0.8 & {mmHg ml$^{-1}$} \\
$E_{min}$ & 0.08 & 0.04 & 0.08 & 0.04 & {mmHg ml$^{-1}$} \\
$m_1$ & 1.32 & 1.32 & 1.32 & 1.32 & {-} \\
$m_2$ & 13.1 & 13.1 & 27.4 & 27.4 & {-} \\
$V_0$ & 3 & 3 & 10 & 10 & {ml} \\
$T_{d}$ & 0.85$T$ & 0.85$T$ & 0 & 0 & {s} \\
\bottomrule
\end{tabular}
}
\end{table}

\vspace{0.5em}  % Adjusts space between tables

\paragraph{Valve} The heart contraction produces a time-varying pressure, and the pressure difference across the valve enforces the valve to open or close. Subsequently, the valve effective area changes during the cardiac cycle (Figure \ref{model_component}(b)). For capturing valve effects on global hemodynamic response, a published 0D valve model \cite{Mynard_Davidson_Penny_Smolich_2012} is used, where the pressure drop across the valve is given by
\begin{equation}
    \Delta P = Bq|q|+L\frac{dq}{dt}.
    \label{valve.eq}
\end{equation}
Here, $B$ is the Bernoulli resistance, $L$ is the blood inertance, and $q$ is blood flow through the valve. The first term represents the turbulence effect of blood, and the second term represents the pressure drag or inertance effect of the blood \cite{Young_Tsai_1973}. The blood flow $q$ is treated as a state variable, and therefore the differential equation for $q$ is obtained by rearranging Equation \eqref{valve.eq} as
\begin{equation} 
    \frac{dq}{dt} = \frac{\Delta P - Bq|q|}{L}.
\end{equation}
The coefficients $B$ and $L$ are dependent on the valve's state:
\begin{equation}
\begin{aligned}
B=\frac{\rho}{2\aeff^{2}}\\
L = \frac{\rho \leff}{\aeff},
\end{aligned}
\end{equation}
where $\rho$ (= 1.06 $g/cm^{3}$ \cite{Mynard_Davidson_Penny_Smolich_2012}) is the blood density, $l_\text{eff}$ and $\aeff$ are respectively effective length and area of the valve. The valve's effective area $\aeff$ varies as
\begin{equation}
\aeff(t) =\aann\xi(t),
\end{equation}
where $\aann$ is valve's annulus area and $\xi$ is the variable that expresses the valve's open/closed status (from 0 to 1, 0 is being completely closed, and 1 is being completely open). Since the valve motion is dominated by the pressure drop across the valve, the change rate of $\xi$ is described as
\begin{equation}
    \frac{d\xi}{dt}=
    \begin{cases} 
    (1-\xi)K_{\text{vo}}\Delta P & \text{if } \Delta P \geq  0 \\
    \xi K_{\text{vc}}\Delta P & \text{if } \Delta P < 0
    \end{cases},
\end{equation}
where $K_\text{vo}$ and $K_\text{vc}$ are the rate coefficient for valve opening and closing. All parameter values for the four valves used in our model are provided in Table \ref{valve_value}.

\begin{table}[htbp]
\caption{Parameter values used for the four heart valves (taken from \cite{Owashi_Hubert_Galli_Donal_Hernandez_Rolle_2020})}
\centering
\sisetup{table-format=1.2} % Adjusts number formatting in siunitx columns
\begin{tabular}{@{}ccccc@{}}
\toprule
& \multicolumn{1}{c}{\textbf{Tricuspid Valve}} & \multicolumn{1}{c}{\textbf{Mitral Valve}} & \multicolumn{1}{c}{\textbf{Aortic Valve}} & \multicolumn{1}{c}{\textbf{Pulmonary Valve}} \\
\midrule
\(l_{\text{eff}}\) (cm) & 2 & 1.25 & 2.2 & 1.9 \\
\(A_{\text{ann}}\) (cm\(^2\)) & 6 & 5 & 5 & 2.8 \\
\(K_{\text{vo}}\) (mmHg\(^{-1}\) s\(^{-1}\)) & 399.9 & 399.9 & 159.9 & 266.6 \\
\(K_{\text{vc}}\) (mmHg\(^{-1}\) s\(^{-1}\)) & 533.3 & 533.3 & 199.9 & 266.6 \\
\bottomrule
\end{tabular}

\label{valve_value}
\end{table}

\paragraph{Circulation} The blood pumped out from the ventricle must overcome the afterload due to the circulatory vasculature \cite{Askari_Messerli_2019, Norton_2001}. Herein, the three-element Windkessel model is adopted to simulate the afterload, one for the systemic circulation and one for the pulmonary circulation. The impedance, peripheral resistance, and compliance are independently represented by $Za$, $R$, and $C$, respectively (Fig. \ref{model_component}(c)). Physiologically, $Za$ and $R$ can be regarded as the resistance of the large and small vessels, respectively \cite{Westerhof_Lankhaar_Westerhof_2009}. If we denote pressure across the capacitor $C$ as $P_{c}$, then the change rate of the $P_{c}$ is given by  
\begin{equation}\label{Lcircuit}
     \frac{dP_{c}}{dt} = \frac{q_{in} - q_{out}}{C} ,
\end{equation}
where $q_{in}$ and $q_{out}$ are the flow rate into and out of the circulation (Figure \ref{model_component}(c)). The proximal pressure is
\begin{equation}\label{prox}
    P_\text{prox} =  Zaq_{in} + P_{c},
\end{equation}
and the distal pressure $P_{\text{distal}}$ is given by 
\begin{equation}\label{distal}
P_{\text{distal}} = P_c - R q_{out}.
\end{equation}

The parameter values of the two Windkessel models remain to be determined, since these parameter values from the literature \cite{Westerhof_Elzinga_Sipkema_1971} produced non-physiological results and therefore were deemed unsuitable for this study. This is likely due to the differences and variations in the model structure of each specific study. The details of parameter value estimation are provided in the section below on \textit{parameter estimation}.

\paragraph{Global response} Equations (\ref{heart chamber}-\ref{distal}), which govern each element, comprise a combination of differential and algebraic equations. These equations are coupled through shared variables, such as pressures at the connections and flows into and out of the elements. To represent the system's dynamics, 14 state variables are selected and organized into the following vector:

\begin{equation}
    \boldsymbol{y} = \left [V_\text{la}, V_\text{lv}, V_\text{ra}, V_\text{rv}, q_\text{mv}, q_\text{av}, q_\text{tv}, q_\text{pv}, \xi_{\text{mv}}, \xi_{\text{av}}, \xi_{\text{tv}}, \xi_{\text{pv}}, P_{1}, P_{2}\right]^\top.
\end{equation}
The combined system of differential equation is written as
\begin{equation}
    \frac{d\boldsymbol{y}}{dt} = \boldsymbol{f}(\boldsymbol{y},t),
\end{equation}
where $\boldsymbol{f}$ is a function based on the Equations (\ref{heart chamber}-\ref{distal}). 
If at any time $t_{n}$, the state variable vector $\boldsymbol{y}_{n}$ is known,  rest of the variables (pressure/ volume/ flow) and the time derivative of state variables at that time can be calculated. Then, the next time state and corresponding state variables can be calculated as:
\begin{equation}\label{derivative}
\begin{aligned}
       t_{n+1} & = t_{n} + \Delta t,\\
       \boldsymbol{y}_{n+1} & = \boldsymbol{y}_{n} + \int\limits_{t_n}^{t_{n+1}} \boldsymbol{f} \,\mathrm{d}t.
\end{aligned}
\end{equation}
The integration in Equation \eqref{derivative} is evaluated using an implicit multi-step variable-order method in Python (Scipy 1.14.1 \cite{2020SciPy-NMeth}). Thus, starting with an assumed initial state at $t=0$, this process is repeated for $N_{iter}$ steps of equal time step $\Delta t$ to obtain the evolution of all pressure/volume/flow variables. The initial values of state variables are listed in Table \ref{initial guess}. Given that the final results showed no significant variation when $\Delta t$ was varied between 0.001 and 0.01, a value of 0.01 was selected to optimise computational efficiency.

\begin{table}[H]
\caption{Initial values for the state variables, time step size, and iteration number \cite{Mynard_Davidson_Penny_Smolich_2012}.}
\centering
\begin{tabular}{lccc}
\hline
\textbf{Variable}         & \textbf{Symbol}   & \textbf{Initial Value} & \textbf{Units} \\ \hline
Left atrial volume        & \(V_\text{la}\)   & 27                    & ml             \\
Left ventricular volume   & \(V_\text{lv}\)   & 135                    & ml             \\
Right atrial volume       & \(V_\text{ra}\)   & 40                    & ml             \\
Right ventricular volume  & \(V_\text{rv}\)   & 180                    & ml             \\
Mitral valve flow         & \(q_\text{mv}\)   & 10                   & ml/s           \\
Aortic valve flow         & \(q_\text{av}\)   & 150                   & ml/s           \\
Tricuspid valve flow      & \(q_\text{tv}\)   & 150                   & ml/s           \\
Pulmonary valve flow      & \(q_\text{pv}\)   & 10                    & ml/s           \\
Mitral valve status    & \(\xi_\text{mv}\) & 0.5                    & -        \\
Aortic valve status    & \(\xi_\text{av}\) & 0.01                    & -        \\
Tricuspid valve status & \(\xi_\text{tv}\) & 0.01                    & -        \\
Pulmonary valve status & \(\xi_\text{pv}\) & 0.5                   & -        \\
Pressure 1                & \(P_1\)           & 5                    & mmHg           \\
Pressure 2                & \(P_2\)           & 5                   & mmHg           \\ \hline
\textbf{Time step size}   & \(\Delta t\)      & 0.01                   & s              \\
\textbf{Iterations}       & \(N_\text{iter}\) & 800                   & -              \\ \hline
\end{tabular}
\label{initial guess}
\end{table}

At the beginning of the simulation, the state variables may exhibit oscillations over a short period. It typically requires several cardiac cycles for the system to reach a steady state. Therefore, 10 cardiac cycles are simulated (for reference, complete waveforms are shown in Figure \ref{waveform}). The key indices (LVEDV, LVESV, RVEDV, RVESV, RVSP, PASP, PADP) from the last cardiac cycle are then post-processed (as the minimum/maximum values, as appropriate) as the model's output. 

\paragraph{Parameter estimation} To find the parameters for two circulations (six in total), an exhaustive search is performed. Using Sobol sampling, a total of 524,288 samples are generated within the specified parameter range (Table \ref{parameter determination}), and the model is evaluated at each of these samples. The parameter ranges were determined based on reference values in the \cite{Westerhof_Elzinga_Sipkema_1971} as benchmarks, on which scaling was performed (e.g., upper and lower limits were scaled up by a factor of 5 each) as a means of assuring that the vast majority of physiologically plausible responses could be covered. Then, the results are compared against physiological reference values of ventricular systolic pressures, stroke volume, ejection fractions, and pulmonary arterial pressure. Specifically, according to \citet{formaggia}, the end-diastolic volume of the left ventricle (LV) ranges from 70–150 mL, the LV end-systolic range is 20–50 mL, the stroke volume is 50–100 mL, and the ejection fraction is 60–80\%. \citet{Broomé_Maksuti_Bjällmark_Frenckner_Janerot-Sjöberg_2013} also summarized the normal human cardiac function. The set of parameters that best aligns with these expected values is selected as the ``baseline'' parameters representing a pre-operative ``healthy'' patient.
\begin{table}[H]
\centering
\caption{The parameter ranges used for the exhaustive search}
\label{parameter determination}
\begin{tabular}{@{}lllllll@{}}
\toprule
      & $C_\text{pul}$   & $Za_\text{pul}$      & $R_\text{pul}$      & $C_\text{sys}$     & $Za_\text{sys}$    & $R_\text{sys}$     \\ \midrule
Range & {[}1.33, 53.33{]} & {[}0.00225, 0.075{]} & {[}0.0225, 0.675{]} & {[}0.2667, 7.99{]} & {[}0.015, 0.45{]}  & {[}0.225,4.5{]}    \\
Unit  & mmHg\(^{-1}\) ml & mmHg ml\(^{-1}\) s   & mmHg ml\(^{-1}\) s  & mmHg\(^{-1}\) ml   & mmHg ml\(^{-1}\) s & mmHg ml\(^{-1}\) s \\ \bottomrule
\end{tabular}
\end{table}
\subsection{Sensitivity analysis}
To investigate how the cardiovascular response is affected by parameter variations, both local and global sensitivity analyses are performed. Since we are primarily interested in the changes in the pulmonary circulation, the effect of parameters $R_\text{pul}$, $Za_\text{pul}$, $C_\text{pul}$, $E_{\max}^{RV}$, and $E_{\min}^{RV}$ is evaluated, as described next.

\subsubsection{Local sensitivity analysis} Local sensitivity is defined as:
\begin{equation}
    \sigma = \frac{\delta y}{y} \left(\frac{\delta x}{x}\right)^{-1},
\end{equation}
where $\sigma$ is the sensitivity value, $y$ is the output of the model, and $x$ is the parameter value. Given that a 1\% perturbation represents a sufficiently small variation, it is adopted as the value for $\frac{\delta x}{x}$. This scale of variation is considered appropriate for local sensitivity analysis, and the resulting sensitivity $\sigma$ is reported as a percentage.

Local sensitivity reveals how much the results change with respect to each parameter, when perturbed independently around the baseline value. However, to gain insights into the relative importance of parameters, when varied simultaneously in a wider range, towards the variation in the results, a global sensitivity analysis is required. 

\subsubsection{Global sensitivity analysis} The global sensitivity analysis of each parameter is carried out following Sobol's approach \cite{Sobol′_2001,Archer_Saltelli_Sobol_1997}, which has two indices that measure the parameter contribution to the variation of a model output: first order effect $S_{1}$ and total effect $S_{T}$. Their expressions for the $i$-th parameter are: 
\begin{equation}\label{sobol}
\begin{aligned}
    S_{1i} &= \frac{D_{i}}{D}\\
     S_{Ti}&=1-\frac{D_{-i}}{D},  
\end{aligned}
\end{equation}
where $D$ is the variance of the specific condition \cite{Saltelli_Annoni_Azzini_Campolongo_Ratto_Tarantola_2010}. $D_{i}$ is the variance caused by $i$-th parameter, and $D_{-i}$ is the variance caused by all parameters, except the ${i}$-th one.
The Sobol sensitivity analysis is implemented using the SALib library in Python \cite{Herman2017, Iwanaga2022}. The parameter range used in this study is provided in Table \ref{sobol_range}. The chosen parameter ranges cover the values under normal conditions and extend to sufficiently cover the majority of the physiological responses. Furthermore, a total of 3,670,016 parameter sets are included in the Sobol sensitivity to guarantee the stability of the analysis.\\
\begin{table}[]
\caption{The parameter ranges used for the global sensitivity analysis}
\centering
\begin{tabular}{@{}llllll@{}}
\toprule
 &
  $C_\text{pul}$ &
  $Za_\text{pul}$ &
  $R_\text{pul}$ &
  $E_{\max}^{RV}$ &
  $E_{\min}^{RV}$ \\ \midrule
Range &
  {[}0.667, 13.333{]} &
  {[}$7.5\times 10^{-4}$, $4.5\times 10^{-2}${]} &
  {[}0.0375, 0.375{]} &
  {[}0.2, 2{]} &
  {[}0.01, 0.16{]} \\
Unit &
  mmHg\(^{-1}\) ml &
  mmHg ml\(^{-1}\) s &
  mmHg ml\(^{-1}\) s &
  mmHg ml$^{-1}$ &
  mmHg ml$^{-1}$ \\ \bottomrule
\end{tabular}
\label{sobol_range}
\end{table}

It should be noted that Sobol indices specifically measure the relative contribution of a parameter to an output's variation. A high Sobol index indicates that a parameter accounts for a large proportion of the output's change, even though the absolute changes in the output may be small. For instance, a parameter might produce very small local variations in the output, yet still yield a high Sobol index because that parameter accounts for majority of the output variations while other parameters do not affect that output. Therefore, to gain a comprehensive understanding of parameter sensitivity, it is important to analyse both local and global sensitivity results simultaneously.

\subsection{Lung Resection Simulation}
As outlined in the preceding section, a gap remains in the modeling of lung resection simulations. To address this gap, a novel model that simulates the cardiovascular system's response following lung resection is introduced.

\subsubsection{Afterload change}
\begin{figure}
    \centering
    \includegraphics[width=1\linewidth]{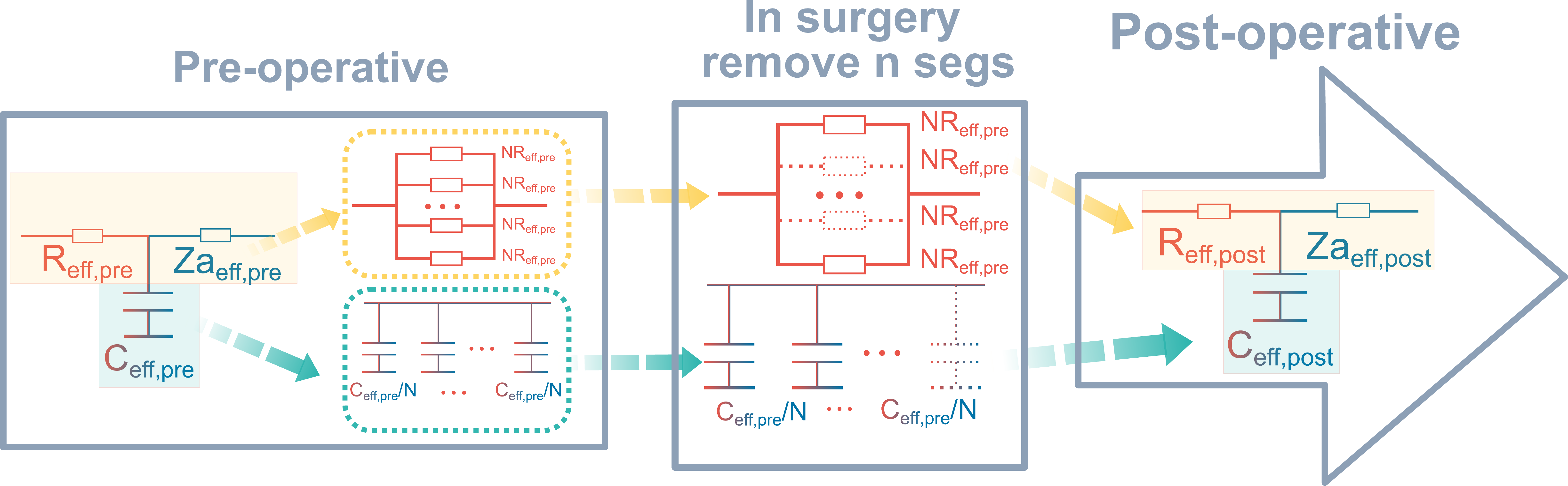}
    \caption{A schematic of the changes in pulmonary circulation during lung resection, categorised into three stages: in the pre-operative stage, the resistance and capacitance of pulmonary circulation are determined from the healthy baseline. Then in surgery implementation stage, $n$ segments are removed, and the effective post-operative values of each element are calculated again at post-operative stage. The expression of insurgery stage is fulfilled through Equation \ref{lung resection formula}-\ref{lung resection formula3} }
    \label{lung_resection_sim}
\end{figure}

Our model for afterload change following lung resection is based on two assumptions:
\begin{itemize}
    \item The lung resection only influences the parameters in the pulmonary circulation
    \item The resistance, impedance, and compliance of each lung segment have the same value at the pre-op stage
\end{itemize}
Based on the majority of reported clinical observations, resistance in the pulmonary circulation is generally considered to increase after surgery. Additionally, due to the reduced number of pulmonary vessels in a physiological context, postoperative pulmonary capacitance is believed to decrease. To conveniently understand, the whole lung resection surgery is divided into three states in the model: pre-op, surgery implementation, and post-op (Figure \ref{lung_resection_sim}). 

\paragraph{Pre-operative status} The pre-operative state is considered to be the same as the healthy baseline condition, and the values of $Za_\text{eff,pre}$, $R_\text{eff,pre}$, and $C_\text{eff,pre}$ determined from the exhaustive search are used.

The actual vessel distribution in the lung is a tree-like structure.  Each pulmonary parameter can be approximated to follow a parallel relationship in the model. Assuming the lung is composed of $N$ segments, and the resistance, impedance, and capacitance of each segment are given by 
\begin{align}
      R_{\text{seg}} &= N R_{\text{eff,pre}},\\
      Za_{\text{seg}} &= N Za_{\text{eff,pre}},\\
      C_{\text{seg}} &= \frac{C_{\text{eff,pre}}}{N},
\end{align}
where subscript $_{\text{eff,pre}}$ represents the effective value under pre-operative conditions.

\paragraph{Surgery implementation} Assume that during the surgical procedure, \(n\) segments are removed, leaving \(N-n\) segments.

\paragraph{Post-operative status} Remaining $N-n$ segments can be combined again to obtain the post-operative values:
\begin{align} 
      R_{\text{eff,post}} &= \frac{N R_{\text{eff,pre}} }{N-n} = \frac{R_{\text{eff,pre}} }{1-\alpha},\label{lung resection formula}\\
      Za_{\text{eff,post}} &= \frac{N Za_{\text{eff,pre}} }{N-n} = \frac{Za_{\text{eff,pre}} }{1-
      \alpha},\\
      C_{\text{eff,post}} &= \frac{(N-n)C_{\text{eff,pre}}}{N} = {(1-\alpha)C_{\text{eff,pre}}},\label{lung resection formula3}
\end{align}\\
where subscript $_{\text{eff,post}}$ represents effective value under post-operative conditions and $\alpha = 100\% \dfrac{n}{N}$ is defined as ratio of the lung volume removed in the surgery. The value of $\alpha$ is varied from 0 to 50\% (clinically plausible range of lobectomy to pneumonectomy) to study the effect of different afterload changes. 

\subsubsection{Contractility change}
Recall Section \ref{intro}, it has been hypothesised that an acute increase in afterload may trigger an inflammatory injury to the RV resulting in impaired function. We assume that the physiological indication of such an RV injury is the reduction of its contractility. In the adopted time-variant elastance model, the maximal right ventricular elastance $E_\text{max}^{RV}$ represents its contractility. Therefore, the reduction of the contractility is seen as the reduction of $E_\text{max}^{RV}$. Thus, for simulating the pathological condition of RV contractility loss, the value of $E_\text{max}^{RV}$ is decreased from the baseline value of 0.8 {mmHg ml$^{-1}$}(Table \ref{heart_chamber_parameter}) to 0.4 {mmHg ml$^{-1}$} (50\% of the initial value) with 1 \% intervals.

\section{Results}
\subsection{Healthy baseline}
\begin{figure}[H]
    \centering
    \includegraphics[width=1.1\linewidth]{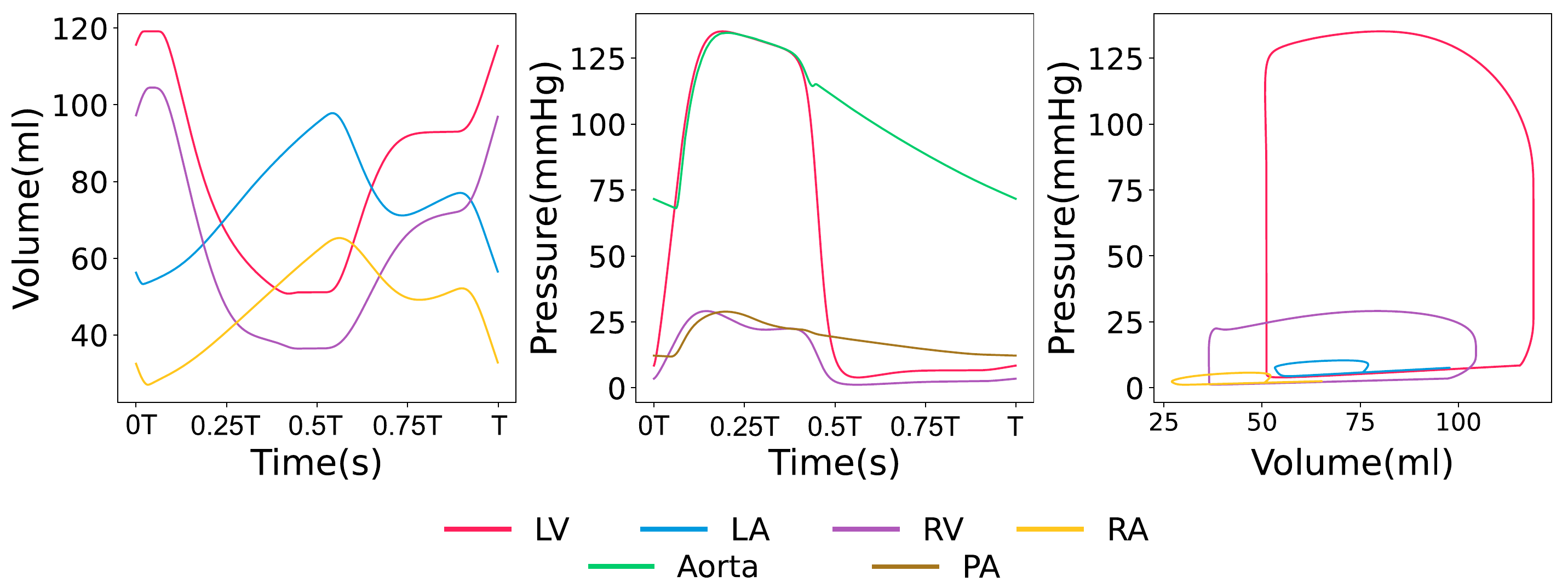}
    \caption{Model results under the baseline condition showing ventricular pressure and volume alongside pressure in the aorta and pulmonary artery.}
    \label{healthy_results}
\end{figure}

 The parameter values of the pulmonary circulation and the systemic circulation obtained using the exhaustive search are given in Table \ref{circulation}. Then, the 0D model presented in Section \ref{model} with these parameters is used to simulate a healthy individual's cardiovascular system, termed as the baseline condition. Figure \ref{healthy_results} shows the baseline results for various cardiovascular pressures and volumes over a cardiac cycle. The pressure variations in ventricles and arteries match the expected trends reported in the literature \cite{Hall_Hall_Guyton_2021, formaggia, Broomé_Maksuti_Bjällmark_Frenckner_Janerot-Sjöberg_2013} and those observed in physiological simulation software such as Circadapt \cite{arts2005adaptation}.   The baseline model outputs of the global measures and corresponding physiologically healthy ranges are listed in Table \ref{ventricle_parameters}. All outputs are reasonably close to the reference values. 
\begin{table}[H]
\caption{Parameter values for systemic and pulmonary circulation determined by exhaustive search.}
\centering
\sisetup{
    table-format=4.5,
    group-digits=false,
    table-space-text-post=\,\textsuperscript{-1}
}
\begin{tabular}{@{} l c l c l @{}}
\toprule
\multicolumn{2}{c}{\textbf{Systemic Circulation}} & \multicolumn{2}{c}{\textbf{Pulmonary Circulation}} & \textbf{Units} \\
\cmidrule(lr){1-2} \cmidrule(lr){3-4} 
\textbf{Parameter} & \textbf{Value} & \textbf{Parameter} & \textbf{Value} & \\
\midrule
\(Za_{\text{sys}}\) & 0.0776 & \(Za_{\text{pul}}\) & 0.01995 & mmHg ml\(^{-1}\) s\\
\(R_{\text{sys}}\) & 1.1032 & \(R_{\text{pul}}\) & 0.1237 &  mmHg ml\(^{-1}\) s \\
\(C_{\text{sys}}\) & 0.8131 & \(C_{\text{pul}}\) & 3.6258 & mmHg\(^{-1}\) ml \\
\bottomrule
\end{tabular}
\label{circulation}
\end{table}

\begin{table}[h]
\caption{Comparison between reference physiological ranges and model results.}
\centering
\setlength{\tabcolsep}{12pt}
\renewcommand{\arraystretch}{1.2}
\begin{tabular}{lcc}
\toprule
\textbf{Parameter} & \textbf{Model} & \textbf{Reference} \\ 
\midrule
RVSV (mL)    & 68.10 & 50–100 \cite{formaggia} \\
RVEF (\%)    & 65.14 & 61$\pm$5.8 \cite{Broomé_Maksuti_Bjällmark_Frenckner_Janerot-Sjöberg_2013} \\
LVEF (\%)    & 57.40 & 59$\pm$4 \cite{Broomé_Maksuti_Bjällmark_Frenckner_Janerot-Sjöberg_2013} \\
LVSP (mmHg)  & 135.18 & 130 \cite{fowler_1980} \\
RVSP (mmHg)  & 29.11 & 27.3$\pm$5.7 \cite{Broomé_Maksuti_Bjällmark_Frenckner_Janerot-Sjöberg_2013} \\
PASP (mmHg)   & 28.9 & 22 \cite{Broomé_Maksuti_Bjällmark_Frenckner_Janerot-Sjöberg_2013} \\
PADP (mmHg)   & 11.85 & 9 \cite{Broomé_Maksuti_Bjällmark_Frenckner_Janerot-Sjöberg_2013} \\
\bottomrule
\end{tabular}
\label{ventricle_parameters}
\end{table}

\subsection{Sensitivity analyses}

\begin{figure}
    \centering
    \includegraphics[width=0.6\linewidth]{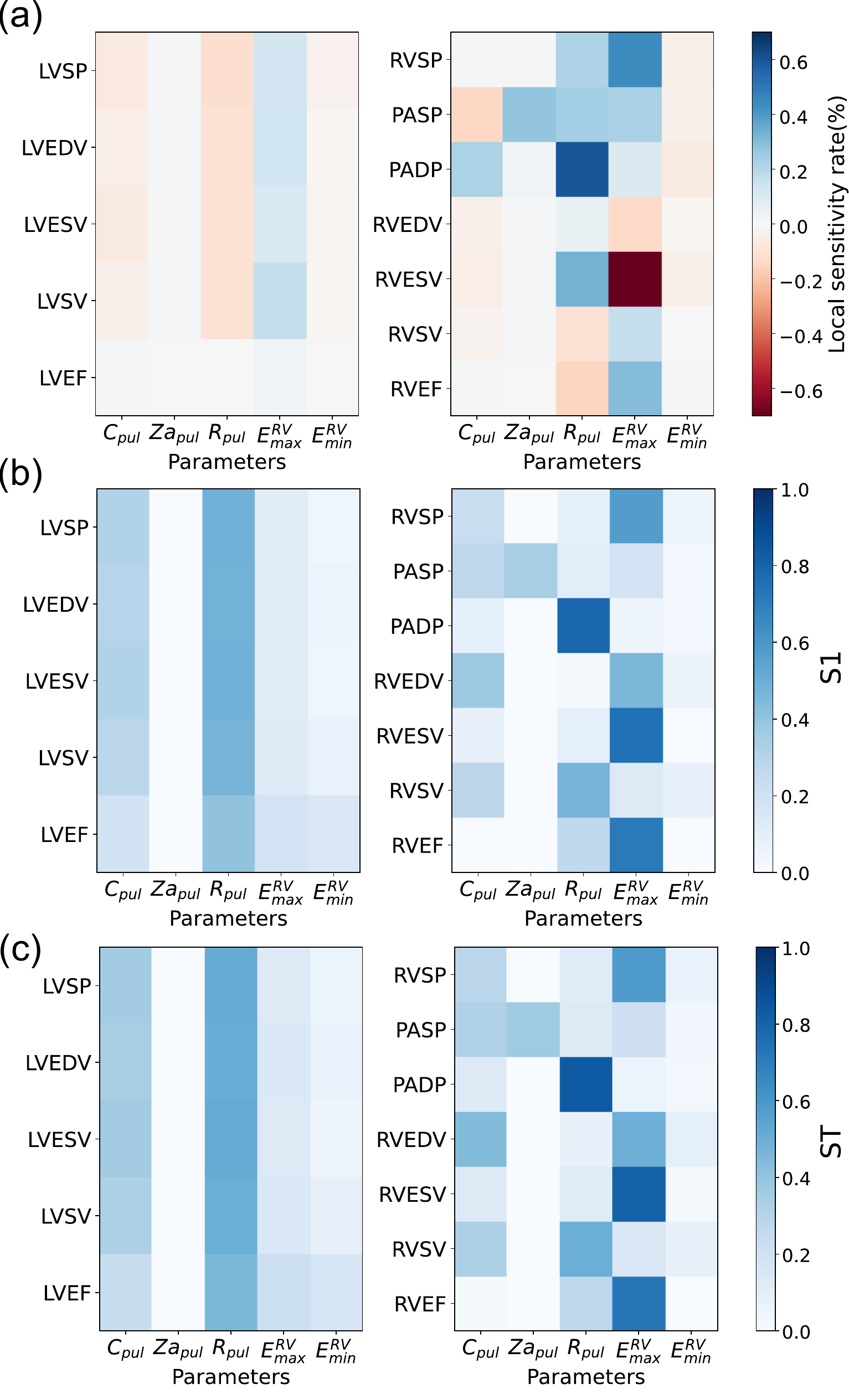}
    \caption{Results of local (a) and global analyses (b, c): (a) local sensitivity shows results change (\%) for 1\% parameter change (b) $S_{1}$ (Equation \ref{sobol}) for Sobol analysis, and (c) $S_{T}$ (Equation \ref{sobol}) for Sobol analysis. Sobol sensitivity results show the contribution of each parameter to the change (a value of 1 indicates complete dependence on the parameter, while a value of 0 signifies no relationship at all). $R_\text{pul}$ and $E_\text{max}^{RV}$ have the highest impact on model results.
}
    \label{sensitivity_analysis}
\end{figure}

The local sensitivity analysis results are displayed in Figure \ref{sensitivity_analysis} (a), and Table \ref{sensitivity_table} shows the top 10 most significant changes in response to parameter variations. The results demonstrate that $R_\text{pul}$ and $E_{\max}^{RV}$ cause the most notable impacts. Specifically, a 1\% change in $R_\text{pul}$ leads to a 0.54\% change in PADP, a 0.3\% change in RVESV, and a 0.19\% change in PASP. Similarly, a 1\% change in $E_{\max}^{RV}$ results in a -0.74\% change in RVESV, a 0.32\% change in RVSP, and a 0.28\% change in RVEF. The results show that the capacitance of the pulmonary circulation $C_\text{pul}$ is positively correlated with PADP and negatively correlated with PASP. The impedance of the pulmonary artery, $Za_\text{pul}$, only influences the PASP and their correlation is positive.

Figure \ref{sensitivity_analysis} (b, c) presents the Sobol sensitivity indices, $S_{1}$ and $S_{T}$. The small difference between $S_{1}$ and $S_{T}$ suggests that the interaction effects among the five parameters ($R_\text{pul}$, $Za_\text{pul}$, $C_\text{pul}$, $E_{\max}^{RV}$, and $E_{\min}^{RV}$) are minimal. Several outputs, such as PADP, RVSV, and LVSV, are highly dependent on $R_\text{pul}$, with a sensitivity index value greater than 0.5. On the other hand, $E_{\max}^{RV}$ predominantly influences RVSP, RVEDV, and RVEF.

It is important to note that Sobol indices can only be compared between different parameters for the same output (i.e., reading a row in Figure \ref{sensitivity_analysis}b,c). That is, one should not compare  $S_1$ of the same parameter on different outputs; for that we should refer to the local sensitivity results. Overall, both local and Sobol sensitivity analyses show that $R_\text{pul}$ and $E_{\max}^{RV}$ are the most influential parameters, followed by $C_{\text{pul}}$. 
\begin{table}[]
\caption{Top 10 absolute local sensitivities and their corresponding parameters and outputs}
\centering
\begin{tabular}{@{}llll@{}}
\toprule
Sorted Rank & Local sensitivity (\%) & Ouput & Parameter           \\ \midrule
1     & -0.74             & RVESV  & $E_\text{max}^{RV}$ \\
2     & 0.54               & PADP   & $R_\text{pul}$      \\
3     & 0.32             & RVSP   & $E_\text{max}^{RV}$ \\
4     & 0.3              & RVESV  & $R_\text{pul}$      \\
5     & 0.28              & RVEF   & $E_\text{max}^{RV}$ \\
6     & 0.27              & PASP   & $Za_\text{pul}$     \\
7     & -0.22              & RVEDV   & $E_\text{max}^{RV}$      \\
8     & -0.21              & PASP   & $C_\text{pul}$ \\
9     & 0.19             & PASP   & $R_\text{pul}$      \\
10    & 0.18              & PADP   & $C_\text{pul}$      \\ \bottomrule
\end{tabular}
\label{sensitivity_table}
\end{table}

\subsection{Effect of lung resection}
\begin{figure}
    \centering
    \includegraphics[width=1\linewidth]{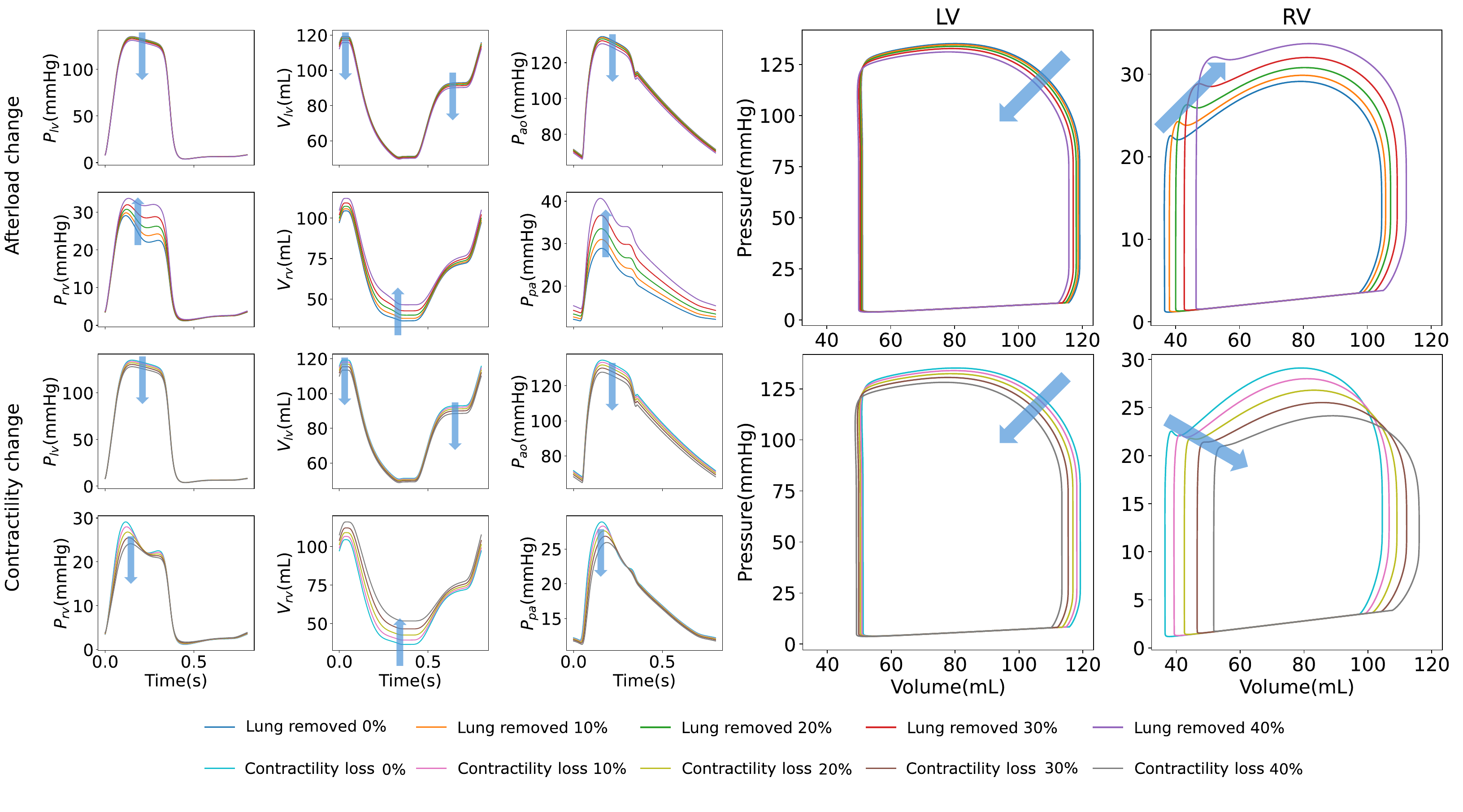}
    \caption{Results of the lung resection simulation under afterload and contractility change cases. Most outputs exhibit similar trend changes in both cases except RVSP, PASP, and PADP. The blue arrow describes the changing trend with the increase of afterload or contractility loss.
}
    \label{resection_change}
\end{figure}

\begin{figure}
    \centering
    \includegraphics[width=1\linewidth]{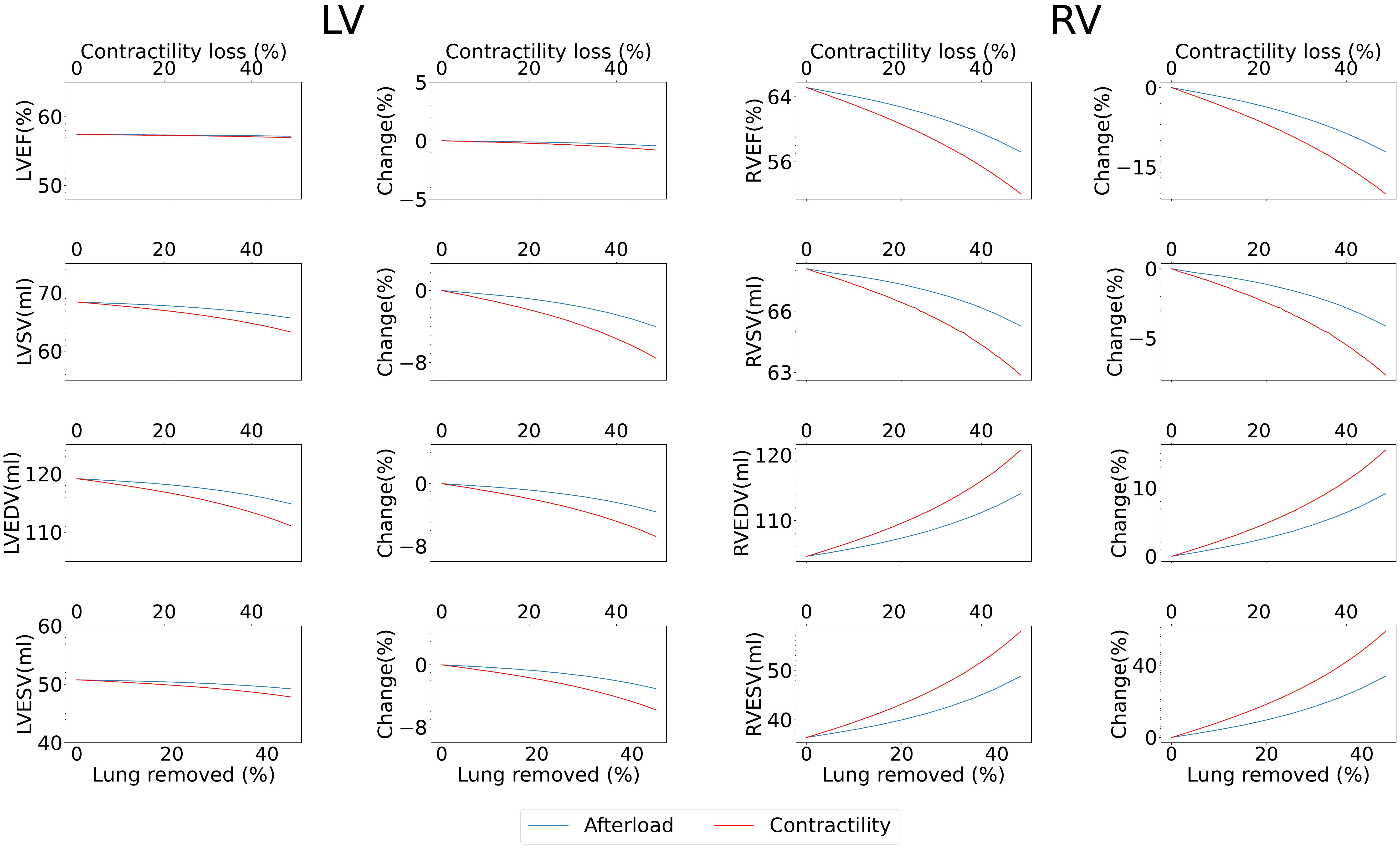}
    \caption{Comparison of the effects of afterload and RV contractility changes on ventricular volume-related indices: all indices in this figure exhibit the same trend under both conditions, with changes in the LV being smaller than those in the RV.}
    \label{volume_remove}
\end{figure}

After lung resection, the afterload and the RV contractility may change. The effect of these changes on the cardiovascular system is simulated, and the resulting pressure and volume waveforms over a cardiac cycle are shown in Figure \ref{resection_change}. Most parameters, including LV pressure ($P_{lv}$), LV volume ($V_{lv}$), and aortic pressure ($P_{ao}$), exhibit similar trends under both afterload and contractility changes. On the LV side, $P_{lv}$, $V_{lv}$, and $P_{ao}$ decrease slightly in both conditions. In contrast, on the RV side, $V_{rv}$ increases significantly (up to about 50\%, Figure \ref{volume_remove}).

There are notable differences in RV pressure ($P_{rv}$) and pulmonary artery pressure ($P_{pa}$) between afterload changes and contractility changes. First, $P_{rv}$ and $P_{pa}$ increase in the afterload-increase scenario, but decrease in the contractility-loss scenario. Second, the timing of these changes differs: with afterload increase, the increases span the entire systolic period (0.1s--0.4s), whereas with contractility loss, the reduction happens mainly in the first half of the systole (0.1s--0.25s).

\paragraph{Comparison of PV-loop}
The pressure-volume (PV) loops exhibit changes with the afterload increase and contractility loss. The LV PV loop exhibits a comparable pattern under both conditions, showing decreases in maximal volume and pressure and shifting toward the lower-left corner. However, the RV PV loop behaves differently: under afterload increase, it moves toward the upper-right corner (indicating higher maximal pressures and volumes), where under contractility loss, it shifts toward the lower-right corner (indicating lower maximal pressures and higher maximal volumes).

\begin{figure}
    \centering
    \includegraphics[width=1\linewidth]{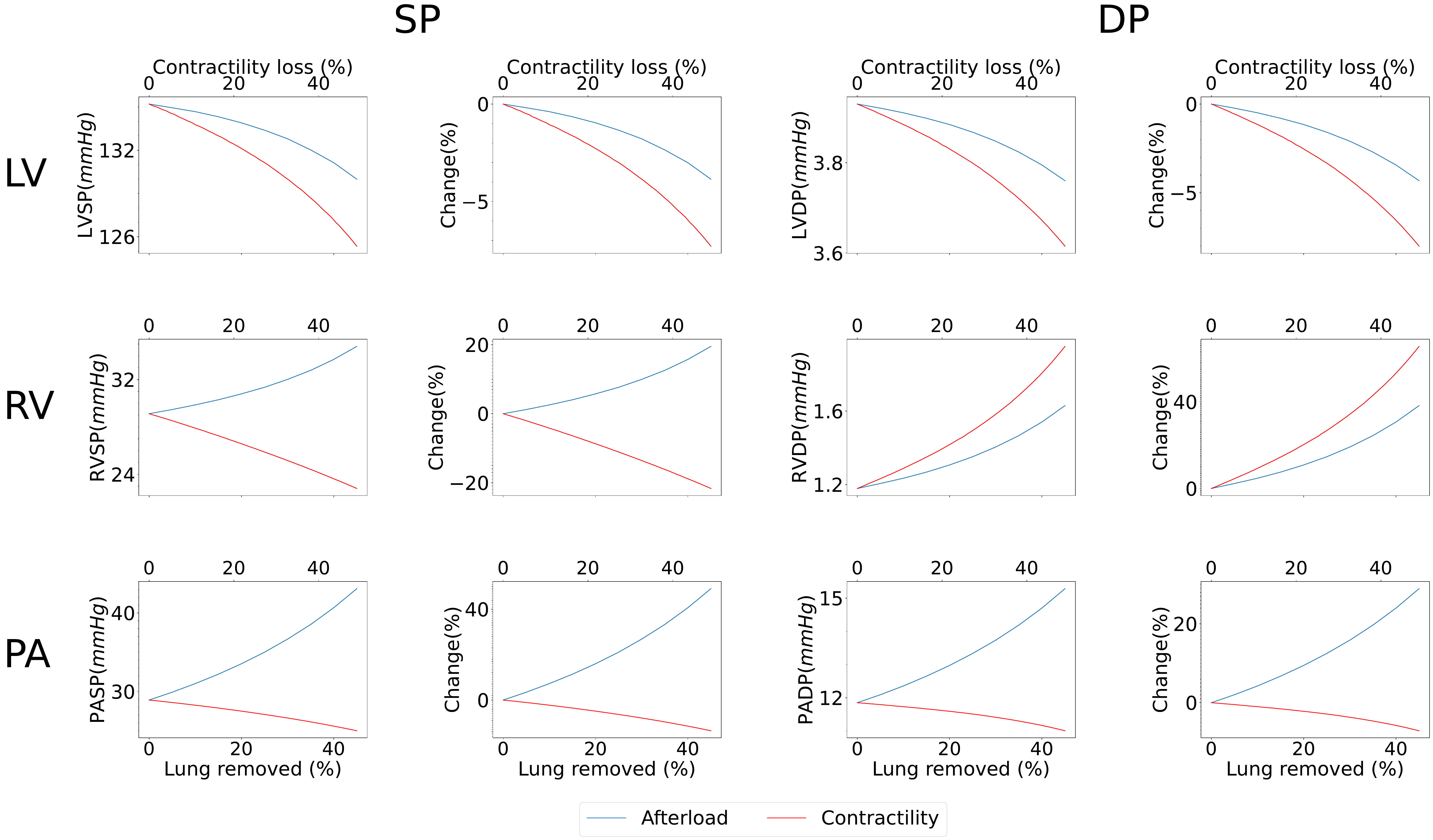}
    \caption{Comparison of the effects of afterload and RV contractility changes on ventricular and pulmonary pressure-related indices: most indices in this figure exhibit the same trend under both conditions except RVSP, PASP, and PADP.}
    \label{pressure_related}
\end{figure}

An important indicator for any patient is the relative change in their cardiovascular function. Thus, \%age-changes in the important systolic/diastolic volumes and pressures from lung resection simulation are presented in Figure \ref{volume_remove} and \ref{pressure_related}.

\paragraph{Systemic System (LV) Response}
The systemic circulation, represented by the LV, experiences minimal impact under both afterload change and altered contractility conditions. When 50\% of the lung is removed, LVEF remains nearly unchanged, with a reduction of less than 1\%, while LVSV, LVEDV, and LVESV decrease slightly (Figure \ref{volume_remove}), each by less than 5\%. The same trend occurs in contractility change cases in LVSV, LVEDV, and LVESV, but the decreasing magnitude is a bit larger than in afterload cases (-7.53\%, -6.78\% and -5.78\% respectively for LVSV, LVEDV, LVESV). Similarly, changes in afterload and contractility lead to modest reductions in LV pressures (Figure \ref{pressure_related}), with LVSP and LVDP decreasing by 3.9\%, 4.3\%, and 7.29\%, 8.01\%, respectively in the two scenarios.

\paragraph{Pulmonary System (RV) Response}
In contrast to systemic circulation, the pulmonary circulation, represented by the RV, undergoes substantial functional alterations. Following a 50\% lung resection, RVEF declines by 12.2\%, and RVSV decreases by 4.1\%. Meanwhile, RVEDV and RVESV expand significantly, increasing by 9.15\% and 34\%, respectively. Similar trends appears under contractility change cases, and all functional changes are larger than the afterload cases. Under elevated afterload, RV pressures rise sharply, with RVSP increasing by 19.5\%, and pulmonary arterial pressures (PASP and PADP) surging by 49.1\% and 28.9\%. However, under contractility reduction, these pulmonary pressures exhibit an opposite trend, with RVSP decreasing by 21.7\%, PASP by 13.5\%, and PADP by 7.12\%, highlighting the distinct hemodynamic responses between systemic and pulmonary systems.
\clearpage

\section{Discussion}
Surgical lung resection remains a first-line treatment for lung cancer patients. Despite successful tumor resection, patients commonly experience functional limitations due to breathlessness and reduced exercise capacity, post-operatively. The extent of functional limitation is not fully explained by the extent of lung tissue removal \cite{Pelletier_Lapointe_LeBlanc_1990, R._Larsen_Svendsen_Milman_Brenøe_Petersen_1997}. Therefore, multiple clinical studies have looked at the effect of lung resection on the cardiovascular system, and more specifically the right heart. Observational clinical studies have consistently shown a reduction in RVEF post-operatively but there remains uncertainty regarding the underlying mechanism. 

This study had two specific aims: 1) to review the published literature of clinical studies reporting the effect of lung resection on the cardiovascular system, and 2) to present the first mathematical model able to describe these effects. The clinical studies summarised in Table \ref{RVdetail} have consistently reported changes in the RV volumes and,  as a consequence, a decrease in the RVEF. The volumes are generally measured using in-vivo imaging such as MRI. The RV and pulmonary artery pressures are usually observed to increase (Table \ref{PAdetail}), with the exception of the findings by \citet{Okada_Ota_Okada_Matsuda_Okada_Ishii_1994}. Most studies found no evidence of significant changes in left heart function, volume or pressure.

\subsection{First computational model}
Even though there is a large amount of literature on the computational modeling of the cardiovascular system, there is no published model of the effects of lung resection on the cardiovascular system. Thus, this study presents the first computational model of the effect of lung resection on the complete closed-loop cardiovascular system---both systematic and pulmonary parts. The model presented here is based on lumped parameter modeling. The focus is on changes in the right heart chambers and pulmonary circulation, and, in particular, the model implements two potential mechanisms behind the effect of lung resection: pulmonary afterload increase and RV contractility loss. 

\subsection{Isolated Parameter Effects}
One advantage of computational models is the ability to study isolated effect of individual components of the closed-loop system. Therefore, this study also investigated cardiovascular responses to individual pulmonary circulatory parameter changes (Figure \ref{sensitivity_analysis}). Results indicated that isolated changes in pulmonary vascular compliance ($C_\text{pul}$) primarily affected PASP and PADP. Changes in large pulmonary vessel impedance ($Za_\text{pul}$) predominantly influenced PASP. In contrast, the capillary resistance ($R_\text{pul}$) variations positively impacted  LVSP, LVEDV, LVESV, and LVSV, while having a highly negative impact on RVSP, PASP, PADP, and RVESV (Figure \ref{sensitivity_analysis}). 

\subsection{Comparison with Previous Clinical Studies}
\citet{ Kowalewski_Brocki_Dryjański_Kaproń_Barcikowski_1999, McCall_Arthur_Glass_Corcoran_Kirk_Macfie_Payne_Johnson_Kinsella_Shelley_2019, Elrakhawy_Alassal_Shaalan_Awad_Sayed_Saffan_2018}, and \citet{Reed_Dorman_Spinale_1993} conducted measurements of RV volumetric parameters (Table \ref{RVdetail}). All clinical studies reported post-operative RVEDV increases, which is consistent with our simulation results---both under afterload increase and RV contractility loss scenarios (Figure \ref{volume_remove}). The study by \citet{McCall_Arthur_Glass_Corcoran_Kirk_Macfie_Payne_Johnson_Kinsella_Shelley_2019} found an increase in RVESV and RVSV, as well as a reduction in RVEF. Similarly, \citet{Reed_Dorman_Spinale_1993, Reed_Spinale_Crawford_1992} also reported a decrease in RVEF. These clinical findings are also consistent with our simulation outcomes under both scenarios. Lastly, previous clinical studies found a small, but non-significant decrease in the LVEF \cite{McCall_Arthur_Glass_Corcoran_Kirk_Macfie_Payne_Johnson_Kinsella_Shelley_2019, Mandal_Dutta_Kumar_Kumar_Ganesan_Bhat_2017}, which is also the prediction from our computational model based on both mechanisms. Thus, based on the commonly used volume measurements, the effects of the two mechanisms are qualitatively indistinguishable. 

In contrast, pressure measurements on the pulmonary side require invasive catheterisation, and are performed less commonly. Moreover, the reported pressure changes in the literature show varying and, sometimes, contradictory trends. \citet{Elrakhawy_Alassal_Shaalan_Awad_Sayed_Saffan_2018} and \citet{Nishimura_Haniuda_Morimoto_Kubo_1993} reported increased mean pulmonary arterial pressure (mPAP), and \citet{Reed_Spinale_Crawford_1992, Reed_Dorman_Spinale_1993} noted increase in all pulmonary pressures (PASP, PADP and mPAP). The only statistically significant change in these studies was the one observed by \citet{Elrakhawy_Alassal_Shaalan_Awad_Sayed_Saffan_2018}. \citet{Amar_Burt_Roistacher_Reinsel_Ginsberg_Wilson_1996} and \citet{Mandal_Dutta_Kumar_Kumar_Ganesan_Bhat_2017} reported increased RVSP, also a statistically insignificant change. However, \citet{Okada_Ota_Okada_Matsuda_Okada_Ishii_1994} reported a slight but statistically non-significant decrease in PASP and mPAP, showing an opposite trend compared to the other findings. The varying trends make it challenging to compare our computational model results with the literature. Notably, under an afterload increase, our model shows increased pulmonary pressures, while under RV contractility loss, our model predicts a decrease in the right pressures. 

\subsection{Role of computational modeling}

Given the relative simplicity of the model presented here, it is remarkable that many of the results align with the clinical observations. It is further reassuring that, in our simulation results, most contractility-related effects were the same as the afterload-related changes, except for RVSP, PASP, and PADP, since these are the expected trends, physiologically.

The underlying mechanisms of postoperative RV dysfunction remain unclear. Many clinical studies attribute post-operative RV dysfunction to afterload changes \cite{Reed_Dorman_Spinale_1993, Amar_Burt_Roistacher_Reinsel_Ginsberg_Wilson_1996, Elrakhawy_Alassal_Shaalan_Awad_Sayed_Saffan_2018}. Our simulations certainly confirm that cardiovascular responses under isolated afterload changes produce trends similar to the clinical observations. 
 However, the similar trends in most volume and pressure parameters between RV contractility and afterload changes suggest that post-op RV dysfunction may result from combined effects of afterload changes and right ventricular contractility changes, rather than afterload effects alone.  In deed, some studies suggest that post-operative RV dysfunction is linked to changes in RV contractility \cite{Mageed_FaragEl-Ghonaimy_Elgamal_Hamza_2005, McCall_Arthur_Glass_Corcoran_Kirk_Macfie_Payne_Johnson_Kinsella_Shelley_2019}. However, the relative extent of the two mechanisms is unknown due to the challenges in clinically measuring the afterload and RV contractility changes after lung resection. 
 
 A computational model is an alternative way to test different hypotheses and thereby complement the clinical data to answer these challenging questions. The first model presented here provides a  starting point for such investigations. The present study is aimed at analysing the \emph{qualitative} trends. A \emph{quantitative} comparison will enable us to determine the relative importance of the two mechanisms. To that end, some of the assumptions underlying our computational model should be further investigated: 1) the model's starting baseline condition is assumed to be ``healthy'', even though the lung cancer patients may have underlying lung and cardiac comorbidity; 2) the model does not account for compensatory/regulatory mechanisms and long-term adaptations; and 3) we assume no coupling between the LV and RV. Moreover, the assumption that pulmonary vascular resistance proportionally relates to lung volume requires validation. Thus, there are several areas where the model presented here will be further developed in the future. 

\subsection{Conclusion}
The first computational model was proposed to simulate the response of cardiovascular system after lung resection. The model implements two potential mechanisms: afterload increase and RV contractility loss. Our simulation results  align with most of the clinical observations, following both mechanisms. These results indicates that RV dysfunction may be caused by both post-op afterload increase and RV contractility loss simultaneously. However, the trends in RVSP, PASP and PADP are opposite in the two scenarios, and, therefore, these could be used to differentiate and identify the underlying reasons. This model opens the door for further development, with the potential to complement the clinical research and thereby uncover the underlying reasons for RV dysfunction following lung resection. 
\section*{Appendix}
\setcounter{subsection}{0}
\renewcommand\thesubsection{A.\arabic{subsection}}
\subsection{Definition of clinical abbreviations}\label{clinical definition}
To avoid the absence of physiological knowledge that influences the understanding of this study, we provide a basic and simple definition, and explain the computational approach for common abbreviations which used in this paper.

\textbf{EDV:} end-diastolic volume, which represents the maximum volume over a cardiac cycle.

\textbf{ESV:} end-systolic volume, which represents the minimum volume over a cardiac cycle.

\textbf{SV:} stroke volume, which represents the blood volume that is pumped by the heart chamber during one cardiac cycle. Thus, it is the difference of the diastolic and systolic volumes, $\text{SV} = \text{EDV}-\text{ESV}$.

\textbf{EF:} ejection fraction, it represents the cardiac function and is calculated as:
\begin{equation}
    \text{EF} = \frac{\text{SV}}{\text{EDV}}.
\end{equation}

\textbf{BSA:} body surface area.

\textbf{Index metrics (EDVI, ESVI, SVI):} the value of the volume metrics after normalizing. These are calculated by dividing by BSA. For example,
\begin{equation}
    \text{EDVI} = \frac{\text{EDV}}{\text{BSA}}
\end{equation}

\textbf{DP:} diastolic pressure, which represents the minimum pressure over a cardiac cycle.

\textbf{SP:} systolic pressure, which represents the maximum pressure over a cardiac cycle.

\subsection{Complete waveform of model pressure and volume}
\renewcommand{\thefigure}{A.\arabic{figure}}
\setcounter{figure}{0}
\begin{figure}[H]
    \centering
    \includegraphics[width=1\linewidth]{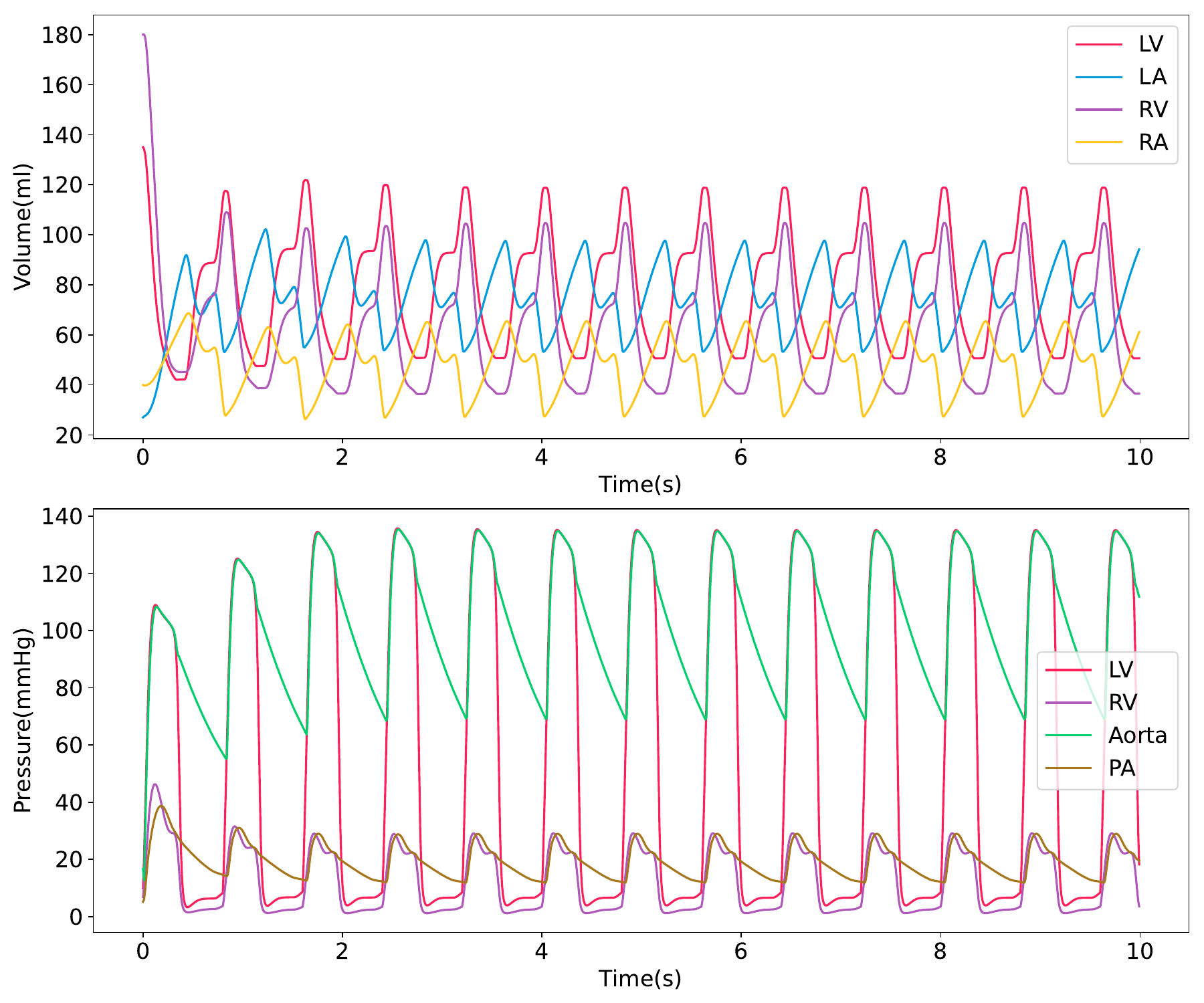}
    \caption{The ten cardiac cycles of pressure and volume waveforms under the baseline condition}
    \label{waveform}
\end{figure}

\clearpage
\bibliography{document}

\begin{thebibliography}{51}
\providecommand{\natexlab}[1]{#1}
\providecommand{\url}[1]{\texttt{#1}}
\expandafter\ifx\csname urlstyle\endcsname\relax
  \providecommand{\doi}[1]{doi: #1}\else
  \providecommand{\doi}{doi: \begingroup \urlstyle{rm}\Url}\fi

\bibitem[Travis(2011)]{Travis_2011}
William~D. Travis.
\newblock Pathology of lung cancer.
\newblock \emph{Clinics in Chest Medicine}, 32\penalty0 (4):\penalty0 669–692, December 2011.
\newblock ISSN 0272-5231, 1557-8216.
\newblock \doi{10.1016/j.ccm.2011.08.005}.

\bibitem[Vainshelboim et~al.(2015)Vainshelboim, Fox, Saute, Sagie, Yehoshua, Fuks, Schneer, and Kramer]{Vainshelboim_Fox_Saute_Sagie_Yehoshua_Fuks_Schneer_Kramer_2015}
Baruch Vainshelboim, Benjamin~Daniel Fox, Milton Saute, Alexander Sagie, Liora Yehoshua, Leonardo Fuks, Sonia Schneer, and Mordechai~R. Kramer.
\newblock Limitations in exercise and functional capacity in long-term postpneumonectomy patients.
\newblock \emph{Journal of Cardiopulmonary Rehabilitation and Prevention}, 35\penalty0 (1):\penalty0 56–64, January 2015.
\newblock ISSN 1932-7501.
\newblock \doi{10.1097/HCR.0000000000000085}.

\bibitem[NICE(2024)]{niceOverviewLung}
NICE.
\newblock {O}verview | {L}ung {C}ancer: diagnosis and management | {G}uidance, 2024.
\newblock URL \url{https://www.nice.org.uk/guidance/ng122}.

\bibitem[McCall et~al.(2019)McCall, Arthur, Glass, Corcoran, Kirk, Macfie, Payne, Johnson, Kinsella, and Shelley]{McCall_Arthur_Glass_Corcoran_Kirk_Macfie_Payne_Johnson_Kinsella_Shelley_2019}
Philip~J. McCall, Alex Arthur, Adam Glass, David~S. Corcoran, Alan Kirk, Alistair Macfie, John Payne, Martin Johnson, John Kinsella, and Benjamin~G. Shelley.
\newblock The right ventricular response to lung resection.
\newblock \emph{The Journal of Thoracic and Cardiovascular Surgery}, 158\penalty0 (2):\penalty0 556--565.e5, August 2019.
\newblock ISSN 00225223.
\newblock \doi{10.1016/j.jtcvs.2019.01.067}.

\bibitem[Pelletier et~al.(1990)Pelletier, Lapointe, and LeBlanc]{Pelletier_Lapointe_LeBlanc_1990}
C~Pelletier, L~Lapointe, and P~LeBlanc.
\newblock Effects of lung resection on pulmonary function and exercise capacity.
\newblock \emph{Thorax}, 45\penalty0 (7):\penalty0 497–502, July 1990.
\newblock ISSN 0040-6376.
\newblock \doi{10.1136/thx.45.7.497}.

\bibitem[R.~Larsen et~al.(1997)R.~Larsen, Svendsen, Milman, Brenøe, and Petersen]{R._Larsen_Svendsen_Milman_Brenøe_Petersen_1997}
Klaus R.~Larsen, Ulrik~G Svendsen, Nils Milman, Jørn Brenøe, and Bruno~N Petersen.
\newblock Cardiopulmonary function at rest and during exercise after resection for bronchial carcinoma.
\newblock \emph{The Annals of Thoracic Surgery}, 64\penalty0 (4):\penalty0 960–964, October 1997.
\newblock ISSN 00034975.
\newblock \doi{10.1016/S0003-4975(97)00635-8}.

\bibitem[Mageed et~al.(2005)Mageed, Farag El-Ghonaimy, Elgamal, and Hamza]{Mageed_FaragEl-Ghonaimy_Elgamal_Hamza_2005}
Nabil~A. Mageed, Yasser~A. Farag El-Ghonaimy, Mohamed-Adel~F. Elgamal, and Usama Hamza.
\newblock Acute effects of lobectomy on right ventricular ejection fraction and mixed venous oxygen saturation.
\newblock \emph{Annals of Saudi Medicine}, 25\penalty0 (6):\penalty0 481–485, November 2005.
\newblock ISSN 0256-4947, 0975-4466.
\newblock \doi{10.5144/0256-4947.2005.481}.

\bibitem[Elrakhawy et~al.(2018)Elrakhawy, Alassal, Shaalan, Awad, Sayed, and Saffan]{Elrakhawy_Alassal_Shaalan_Awad_Sayed_Saffan_2018}
Hany~M. Elrakhawy, Mohamed~A. Alassal, Ayman~M. Shaalan, Ahmed~A. Awad, Sameh Sayed, and Mohammad~M. Saffan.
\newblock Impact of major pulmonary resections on right ventricular function: Early postoperative changes.
\newblock \emph{The Heart Surgery Forum}, 21\penalty0 (11):\penalty0 E009--E017, January 2018.
\newblock ISSN 1522-6662.
\newblock \doi{10.1532/hsf.1864}.

\bibitem[Okada et~al.(1994)Okada, Ota, Okada, Matsuda, Okada, and Ishii]{Okada_Ota_Okada_Matsuda_Okada_Ishii_1994}
Morihito Okada, Toshiaki Ota, Masayoshi Okada, Hitoshi Matsuda, Kenji Okada, and Noboru Ishii.
\newblock Right ventricular dysfunction after major pulmonary resection.
\newblock \emph{The Journal of Thoracic and Cardiovascular Surgery}, 108\penalty0 (3):\penalty0 503–511, September 1994.
\newblock ISSN 0022-5223.
\newblock \doi{10.1016/S0022-5223(94)70260-8}.

\bibitem[Kowalewski et~al.(1999)Kowalewski, Brocki, Dryjański, Kaproń, and Barcikowski]{Kowalewski_Brocki_Dryjański_Kaproń_Barcikowski_1999}
Janusz Kowalewski, Marian Brocki, Tadeusz Dryjański, Krzysztof Kaproń, and Stanisław Barcikowski.
\newblock Right ventricular morphology and function after pulmonary resection1.
\newblock \emph{European Journal of Cardio-Thoracic Surgery}, 15\penalty0 (4):\penalty0 444–448, April 1999.
\newblock ISSN 1010-7940.
\newblock \doi{10.1016/S1010-7940(99)00032-9}.

\bibitem[Gelzinis et~al.(2020)Gelzinis, Assaad, and Perrino]{Gelzinis_Assaad_Perrino_2020}
Theresa Gelzinis, Sherif Assaad, and Albert~C. Perrino.
\newblock Right ventricular function during and after thoracic surgery.
\newblock \emph{Current Opinion in Anaesthesiology}, 33\penalty0 (1):\penalty0 27–36, February 2020.
\newblock ISSN 0952-7907, 1473-6500.
\newblock \doi{10.1097/ACO.0000000000000809}.

\bibitem[Shelley et~al.(2023)Shelley, Glass, Keast, McErlane, Hughes, Lafferty, Marczin, and McCall]{Shelley_Glass_Keast_McErlane_Hughes_Lafferty_Marczin_McCall_2023}
Ben Shelley, Adam Glass, Thomas Keast, James McErlane, Cara Hughes, Brian Lafferty, Nandor Marczin, and Philip McCall.
\newblock Perioperative cardiovascular pathophysiology in patients undergoing lung resection surgery: a narrative review.
\newblock \emph{British Journal of Anaesthesia}, 130\penalty0 (1):\penalty0 e66–e79, January 2023.
\newblock ISSN 00070912.
\newblock \doi{10.1016/j.bja.2022.06.035}.

\bibitem[Greyson et~al.(1997)Greyson, Xu, Cohen, and G.~Schwartz]{Greyson_Xu_Cohen_G._Schwartz_1997}
Clifford Greyson, Ya~Xu, Joshua Cohen, and Gregory G.~Schwartz.
\newblock Right ventricular dysfunction persists following brief right ventricular pressure overload.
\newblock \emph{Cardiovascular Research}, 34\penalty0 (2):\penalty0 281–288, May 1997.
\newblock ISSN 0008-6363.
\newblock \doi{10.1016/S0008-6363(97)00038-2}.

\bibitem[Watts et~al.(2010)Watts, Marchick, and Kline]{Watts_Marchick_Kline_2010}
John~A. Watts, Michael~R. Marchick, and Jeffrey~A. Kline.
\newblock Right ventricular heart failure from pulmonary embolism: Key distinctions from chronic pulmonary hypertension.
\newblock \emph{Journal of Cardiac Failure}, 16\penalty0 (3):\penalty0 250–259, March 2010.
\newblock ISSN 1071-9164.
\newblock \doi{10.1016/j.cardfail.2009.11.008}.

\bibitem[Watts et~al.(2009)Watts, Gellar, Stuart, Obraztsova, and Kline]{Watts_Gellar_Stuart_Obraztsova_Kline_2009}
J.A. Watts, M.A. Gellar, L.K. Stuart, M.~Obraztsova, and J.A. Kline.
\newblock Proinflammatory events in right ventricular damage during pulmonary embolism: Effects of treatment with ketorolac in rats.
\newblock \emph{Journal of Cardiovascular Pharmacology}, 54\penalty0 (3):\penalty0 246–252, 2009.
\newblock \doi{10.1097/FJC.0b013e3181b2b699}.

\bibitem[Reed et~al.(1993)Reed, Dorman, and Spinale]{Reed_Dorman_Spinale_1993}
Carolyn~E. Reed, B.~Hugh Dorman, and Francis~G. Spinale.
\newblock Assessment of right ventricular contractile performance after pulmonary resection.
\newblock \emph{The Annals of Thoracic Surgery}, 56\penalty0 (3):\penalty0 426–432, September 1993.
\newblock ISSN 00034975.
\newblock \doi{10.1016/0003-4975(93)90874-H}.

\bibitem[Reed et~al.(1996)Reed, Dorman, and Spinale]{Reed_Dorman_Spinale_1996}
Carolyn~E. Reed, B.~Hugh Dorman, and Francis~G. Spinale.
\newblock Mechanisms of right ventricular dysfunction after pulmonary resection.
\newblock \emph{The Annals of Thoracic Surgery}, 62\penalty0 (1):\penalty0 225–232, July 1996.
\newblock ISSN 0003-4975.
\newblock \doi{10.1016/0003-4975(96)00258-5}.

\bibitem[Rams et~al.(1962)Rams, Harrison, Fry, Moulder, and Adams]{Rams_Harrison_Fry_Moulder_Adams_1962}
James~J. Rams, Robert~W. Harrison, Willard~A. Fry, Peter~V. Moulder, and William~E. Adams.
\newblock Operative pulmonary artery pressure measurements as a guide to postoperative management and prognosis following pneumonectomy.
\newblock \emph{Diseases of the Chest}, 41\penalty0 (1):\penalty0 85–90, January 1962.
\newblock ISSN 00960217.
\newblock \doi{10.1378/chest.41.1.85}.

\bibitem[Reed et~al.(1992)Reed, Spinale, and Crawford]{Reed_Spinale_Crawford_1992}
Carolyn~E. Reed, Francis~G. Spinale, and Fred~A. Crawford.
\newblock Effect of pulmonary resection on right ventricular function.
\newblock \emph{The Annals of Thoracic Surgery}, 53\penalty0 (4):\penalty0 578–582, April 1992.
\newblock ISSN 00034975.
\newblock \doi{10.1016/0003-4975(92)90314-T}.

\bibitem[Okada et~al.(1996)Okada, Okada, Ishii, Yamashita, Sugimoto, Okada, Yamagishi, Yamashita, and Matsuda]{Okada_Okada_Ishii_Yamashita_Sugimoto_Okada_Yamagishi_Yamashita_Matsuda_1996}
Morihito Okada, Masayoshi Okada, Noboru Ishii, Chojiro Yamashita, Takaki Sugimoto, Kenji Okada, Hiroyuki Yamagishi, Teruo Yamashita, and Hitoshi Matsuda.
\newblock Right ventricular ejection fraction in the preoperative risk evaluation of candidates for pulmonary resection.
\newblock \emph{The Journal of Thoracic and Cardiovascular Surgery}, 112\penalty0 (2):\penalty0 364–370, August 1996.
\newblock ISSN 0022-5223, 1097-685X.
\newblock \doi{10.1016/S0022-5223(96)70263-5}.

\bibitem[Nishimura et~al.(1993)Nishimura, Haniuda, Morimoto, and Kubo]{Nishimura_Haniuda_Morimoto_Kubo_1993}
Hideki Nishimura, Masayuki Haniuda, Masami Morimoto, and Keishi Kubo.
\newblock Cardiopulmonary function after pulmonary lobectomy in patients with lung cancer.
\newblock \emph{Ann Thorac Surg}, 1993.

\bibitem[Glass et~al.(2023)Glass, McCall, Arthur, Mangion, and Shelley]{Glass_McCall_Arthur_Mangion_Shelley_2023}
Adam Glass, Philip McCall, Alex Arthur, Kenneth Mangion, and Ben Shelley.
\newblock Pulmonary artery wave reflection and right ventricular function after lung resection.
\newblock \emph{British Journal of Anaesthesia}, 130\penalty0 (1):\penalty0 e128–e136, January 2023.
\newblock ISSN 0007-0912.
\newblock \doi{10.1016/j.bja.2022.07.052}.

\bibitem[Rauch et~al.(2017)Rauch, Marinova, Schild, and Strunk]{Rauch_Marinova_Schild_Strunk_2017}
Maximilian Rauch, Milka Marinova, Hans~Heinz Schild, and Holger Strunk.
\newblock Cardiovascular computed tomography findings after pneumonectomy.
\newblock \emph{Academic Radiology}, 24\penalty0 (7):\penalty0 860–866, July 2017.
\newblock ISSN 10766332.
\newblock \doi{10.1016/j.acra.2017.01.020}.

\bibitem[Amar et~al.(1996)Amar, Burt, Roistacher, Reinsel, Ginsberg, and Wilson]{Amar_Burt_Roistacher_Reinsel_Ginsberg_Wilson_1996}
David Amar, Michael~E. Burt, Nancy Roistacher, Ruth~A. Reinsel, Robert~J. Ginsberg, and Roger~S. Wilson.
\newblock Value of perioperative doppler echocardiography in patients undergoing major lung resection.
\newblock \emph{The Annals of Thoracic Surgery}, 61\penalty0 (2):\penalty0 516–520, February 1996.
\newblock ISSN 0003-4975.
\newblock \doi{10.1016/0003-4975(95)00939-6}.

\bibitem[Mandal et~al.(2017)Mandal, Dutta, Kumar, Kumar, Ganesan, and Bhat]{Mandal_Dutta_Kumar_Kumar_Ganesan_Bhat_2017}
Banashree Mandal, Vikas Dutta, Balbir Kumar, Alok Kumar, Rajarajan Ganesan, and Imran~H Bhat.
\newblock Echocardiographic evaluation of right ventricular function in the immediate postoperative period after major pulmonary resections: A prospective observational study.
\newblock \emph{Journal of Perioperative Echocardiography}, 5\penalty0 (2):\penalty0 42–48, December 2017.
\newblock ISSN 2320-527X, 2320-7310.
\newblock \doi{10.5005/jp-journals-10034-1070}.

\bibitem[Korakianitis and Shi(2006)]{Korakianitis_Shi_2006}
Theodosios Korakianitis and Yubing Shi.
\newblock A concentrated parameter model for the human cardiovascular system including heart valve dynamics and atrioventricular interaction.
\newblock \emph{Medical Engineering \& Physics}, 28\penalty0 (7):\penalty0 613–628, September 2006.
\newblock ISSN 13504533.
\newblock \doi{10.1016/j.medengphy.2005.10.004}.

\bibitem[Shi et~al.(2011)Shi, Lawford, and Hose]{Shi_Lawford_Hose_2011}
Yubing Shi, Patricia Lawford, and Rodney Hose.
\newblock Review of zero-d and 1-d models of blood flow in the cardiovascular system.
\newblock \emph{BioMedical Engineering OnLine}, 10\penalty0 (1):\penalty0 33, April 2011.
\newblock ISSN 1475-925X.
\newblock \doi{10.1186/1475-925X-10-33}.

\bibitem[Mescher and Junqueira(2018)]{Mescher_Junqueira_2018}
Anthony~L. Mescher and Luiz Carlos~Uchôa Junqueira.
\newblock \emph{Junqueira’s basic histology: text and atlas}.
\newblock Mcgraw-Hill Education, New York, fifteenth edition edition, 2018.
\newblock ISBN 978-1-260-02618-4.

\bibitem[Olufsen and Nadim()]{Olufsen_Nadim}
Mette~S Olufsen and Ali Nadim.
\newblock On deriving lumped models for blood flow and pressure in the systemic arteries.

\bibitem[Suga et~al.(1973)Suga, Sagawa, and Shoukas]{Suga_Sagawa_Shoukas_1973}
Hiroyuki Suga, Kiichi Sagawa, and Artin~A. Shoukas.
\newblock Load independence of the instantaneous pressure-volume ratio of the canine left ventricle and effects of epinephrine and heart rate on the ratio.
\newblock \emph{Circulation Research}, 32\penalty0 (3):\penalty0 314–322, March 1973.
\newblock \doi{10.1161/01.RES.32.3.314}.

\bibitem[Suga and Sagawa(1974)]{Suga_Sagawa_1974}
Hiroyuki Suga and Kiichi Sagawa.
\newblock Instantaneous pressure-volume relationships and their ratio in the excised, supported canine left ventricle.
\newblock \emph{Circulation Research}, 35\penalty0 (1):\penalty0 117–126, July 1974.
\newblock ISSN 0009-7330, 1524-4571.
\newblock \doi{10.1161/01.RES.35.1.117}.

\bibitem[Mynard et~al.(2012)Mynard, Davidson, Penny, and Smolich]{Mynard_Davidson_Penny_Smolich_2012}
J.~P. Mynard, M.~R. Davidson, D.~J. Penny, and J.~J. Smolich.
\newblock A simple, versatile valve model for use in lumped parameter and one‐dimensional cardiovascular models.
\newblock \emph{International Journal for Numerical Methods in Biomedical Engineering}, 28\penalty0 (6–7):\penalty0 626–641, June 2012.
\newblock ISSN 2040-7939, 2040-7947.
\newblock \doi{10.1002/cnm.1466}.

\bibitem[Stergiopulos et~al.(1996)Stergiopulos, Meister, and Westerhof]{Stergiopulos_Meister_Westerhof_1996}
N.~Stergiopulos, J.~J. Meister, and N.~Westerhof.
\newblock Determinants of stroke volume and systolic and diastolic aortic pressure.
\newblock \emph{American Journal of Physiology-Heart and Circulatory Physiology}, 270\penalty0 (6):\penalty0 H2050–H2059, June 1996.
\newblock ISSN 0363-6135, 1522-1539.
\newblock \doi{10.1152/ajpheart.1996.270.6.H2050}.

\bibitem[Chung et~al.(1997)Chung, Niranjan, Clark, Bidani, Johnston, Zwischenberger, and Traber]{Chung_Niranjan_Clark_Bidani_Johnston_Zwischenberger_Traber_1997}
D.~C. Chung, S.~C. Niranjan, J.~W. Clark, A.~Bidani, W.~E. Johnston, J.~B. Zwischenberger, and D.~L. Traber.
\newblock A dynamic model of ventricular interaction and pericardial influence.
\newblock \emph{American Journal of Physiology-Heart and Circulatory Physiology}, 272\penalty0 (6):\penalty0 H2942–H2962, June 1997.
\newblock ISSN 0363-6135, 1522-1539.
\newblock \doi{10.1152/ajpheart.1997.272.6.H2942}.

\bibitem[Young and Tsai(1973)]{Young_Tsai_1973}
Donald~F. Young and Frank~Y. Tsai.
\newblock Flow characteristics in models of arterial stenoses — ii. unsteady flow.
\newblock \emph{Journal of Biomechanics}, 6\penalty0 (5):\penalty0 547–559, September 1973.
\newblock ISSN 0021-9290.
\newblock \doi{10.1016/0021-9290(73)90012-2}.

\bibitem[Owashi et~al.(2020)Owashi, Hubert, Galli, Donal, Hernandez, and Rolle]{Owashi_Hubert_Galli_Donal_Hernandez_Rolle_2020}
Kimi~P. Owashi, Arnaud Hubert, Elena Galli, Erwan Donal, Alfredo~I. Hernandez, and Virginie~Le Rolle.
\newblock Model-based estimation of left ventricular pressure and myocardial work in aortic stenosis.
\newblock \emph{PLOS ONE}, 15\penalty0 (3):\penalty0 e0229609, mar 2020.
\newblock ISSN 1932-6203.
\newblock \doi{10.1371/journal.pone.0229609}.

\bibitem[Ask(2019)]{Askari_Messerli_2019}
\emph{Cardiovascular Hemodynamics: An Introductory Guide}.
\newblock Contemporary Cardiology. Springer International Publishing, Cham, 2019.
\newblock ISBN 978-3-030-19130-6.
\newblock \doi{10.1007/978-3-030-19131-3}.
\newblock URL \url{http://link.springer.com/10.1007/978-3-030-19131-3}.

\bibitem[Norton(2001)]{Norton_2001}
James~M. Norton.
\newblock Toward consistent definitions for preload and afterload.
\newblock \emph{Advances in Physiology Education}, 25\penalty0 (1):\penalty0 53–61, March 2001.
\newblock ISSN 1043-4046, 1522-1229.
\newblock \doi{10.1152/advances.2001.25.1.53}.

\bibitem[Westerhof et~al.(2009)Westerhof, Lankhaar, and Westerhof]{Westerhof_Lankhaar_Westerhof_2009}
Nico Westerhof, Jan-Willem Lankhaar, and Berend~E. Westerhof.
\newblock The arterial windkessel.
\newblock \emph{Medical \& Biological Engineering \& Computing}, 47\penalty0 (2):\penalty0 131–141, February 2009.
\newblock ISSN 0140-0118, 1741-0444.
\newblock \doi{10.1007/s11517-008-0359-2}.

\bibitem[Westerhof et~al.(1971)Westerhof, Elzinga, and Sipkema]{Westerhof_Elzinga_Sipkema_1971}
N~Westerhof, G~Elzinga, and P~Sipkema.
\newblock An artificial arterial system for pumping hearts.
\newblock \emph{Journal of Applied Physiology}, 31\penalty0 (5):\penalty0 776–781, November 1971.
\newblock ISSN 8750-7587, 1522-1601.
\newblock \doi{10.1152/jappl.1971.31.5.776}.

\bibitem[Virtanen et~al.(2020)Virtanen, Gommers, Oliphant, Haberland, Reddy, Cournapeau, Burovski, Peterson, Weckesser, Bright, {van der Walt}, Brett, Wilson, Millman, Mayorov, Nelson, Jones, Kern, Larson, Carey, Polat, Feng, Moore, {VanderPlas}, Laxalde, Perktold, Cimrman, Henriksen, Quintero, Harris, Archibald, Ribeiro, Pedregosa, {van Mulbregt}, and {SciPy 1.0 Contributors}]{2020SciPy-NMeth}
Pauli Virtanen, Ralf Gommers, Travis~E. Oliphant, Matt Haberland, Tyler Reddy, David Cournapeau, Evgeni Burovski, Pearu Peterson, Warren Weckesser, Jonathan Bright, St{\'e}fan~J. {van der Walt}, Matthew Brett, Joshua Wilson, K.~Jarrod Millman, Nikolay Mayorov, Andrew R.~J. Nelson, Eric Jones, Robert Kern, Eric Larson, C~J Carey, {\.I}lhan Polat, Yu~Feng, Eric~W. Moore, Jake {VanderPlas}, Denis Laxalde, Josef Perktold, Robert Cimrman, Ian Henriksen, E.~A. Quintero, Charles~R. Harris, Anne~M. Archibald, Ant{\^o}nio~H. Ribeiro, Fabian Pedregosa, Paul {van Mulbregt}, and {SciPy 1.0 Contributors}.
\newblock {{SciPy} 1.0: Fundamental Algorithms for Scientific Computing in Python}.
\newblock \emph{Nature Methods}, 17:\penalty0 261--272, 2020.
\newblock \doi{10.1038/s41592-019-0686-2}.

\bibitem[Formaggia et~al.(2009)Formaggia, Quarteroni, and Veneziani]{formaggia}
Luca Formaggia, Alfio Quarteroni, and Alessandro Veneziani.
\newblock \emph{Cardiovascular Mathematics: Modeling and Simulation of the Circulatory System}, volume~1.
\newblock 01 2009.
\newblock ISBN 978-88-470-1151-9.
\newblock \doi{10.1007/978-88-470-1152-6}.

\bibitem[Broomé et~al.(2013)Broomé, Maksuti, Bjällmark, Frenckner, and Janerot-Sjöberg]{Broomé_Maksuti_Bjällmark_Frenckner_Janerot-Sjöberg_2013}
Michael Broomé, Elira Maksuti, Anna Bjällmark, Björn Frenckner, and Birgitta Janerot-Sjöberg.
\newblock Closed-loop real-time simulation model of hemodynamics and oxygen transport in the cardiovascular system.
\newblock \emph{BioMedical Engineering OnLine}, 12\penalty0 (1):\penalty0 69, July 2013.
\newblock ISSN 1475-925X.
\newblock \doi{10.1186/1475-925X-12-69}.

\bibitem[Sobol(2001)]{Sobol′_2001}
I.M Sobol.
\newblock Global sensitivity indices for nonlinear mathematical models and their monte carlo estimates.
\newblock \emph{Mathematics and Computers in Simulation}, 55\penalty0 (1–3):\penalty0 271–280, February 2001.
\newblock ISSN 03784754.
\newblock \doi{10.1016/S0378-4754(00)00270-6}.

\bibitem[Archer et~al.(1997)Archer, Saltelli, and Sobol]{Archer_Saltelli_Sobol_1997}
G.~E.~B. Archer, A.~Saltelli, and I.~M. Sobol.
\newblock Sensitivity measures,anova-like techniques and the use of bootstrap.
\newblock \emph{Journal of Statistical Computation and Simulation}, 58\penalty0 (2):\penalty0 99–120, May 1997.
\newblock ISSN 0094-9655, 1563-5163.
\newblock \doi{10.1080/00949659708811825}.

\bibitem[Saltelli et~al.(2010)Saltelli, Annoni, Azzini, Campolongo, Ratto, and Tarantola]{Saltelli_Annoni_Azzini_Campolongo_Ratto_Tarantola_2010}
Andrea Saltelli, Paola Annoni, Ivano Azzini, Francesca Campolongo, Marco Ratto, and Stefano Tarantola.
\newblock Variance based sensitivity analysis of model output. design and estimator for the total sensitivity index.
\newblock \emph{Computer Physics Communications}, 181\penalty0 (2):\penalty0 259–270, February 2010.
\newblock ISSN 00104655.
\newblock \doi{10.1016/j.cpc.2009.09.018}.

\bibitem[Herman and Usher(2017)]{Herman2017}
Jon Herman and Will Usher.
\newblock {SALib}: An open-source python library for sensitivity analysis.
\newblock \emph{The Journal of Open Source Software}, 2\penalty0 (9), jan 2017.
\newblock \doi{10.21105/joss.00097}.
\newblock URL \url{https://doi.org/10.21105/joss.00097}.

\bibitem[Iwanaga et~al.(2022)Iwanaga, Usher, and Herman]{Iwanaga2022}
Takuya Iwanaga, William Usher, and Jonathan Herman.
\newblock Toward {SALib} 2.0: {Advancing} the accessibility and interpretability of global sensitivity analyses.
\newblock \emph{Socio-Environmental Systems Modelling}, 4:\penalty0 18155, May 2022.
\newblock \doi{10.18174/sesmo.18155}.
\newblock URL \url{https://sesmo.org/article/view/18155}.

\bibitem[Hall et~al.(2021)Hall, Hall, and Guyton]{Hall_Hall_Guyton_2021}
John~E. Hall, Michael~E. Hall, and Arthur~C. Guyton.
\newblock \emph{Guyton and Hall textbook of medical physiology}.
\newblock Elsevier, Philadelphia, PA, 14th edition edition, 2021.
\newblock ISBN 978-0-323-59712-8.

\bibitem[Arts et~al.(2005)Arts, Delhaas, Bovendeerd, Verbeek, and Prinzen]{arts2005adaptation}
Theo Arts, Tammo Delhaas, Peter Bovendeerd, Xander Verbeek, and Frits~W Prinzen.
\newblock Adaptation to mechanical load determines shape and properties of heart and circulation: the circadapt model.
\newblock \emph{American Journal of Physiology-Heart and Circulatory Physiology}, 288\penalty0 (4):\penalty0 H1943--H1954, 2005.

\bibitem[Fowler(1980)]{fowler_1980}
Noble~O Fowler.
\newblock \emph{Cardiac Diagnosis and Treatment}.
\newblock HarperCollins Publishers, 1980.

\end{thebibliography}
\end{document}